\documentclass[3p,12pt]{elsarticle}
\newtheorem{theorem}{Theorem}
\newtheorem{corollary}{Corollary}

\newtheorem{remark}{Remark}

\newtheorem{example}{Example}
\usepackage{graphics}
\usepackage{amsmath}
\usepackage{amssymb}
\usepackage{epsfig}
\usepackage{epstopdf}
\usepackage{color}
\usepackage[%
breaklinks%
,colorlinks%
,linkcolor=red% internal document links
,anchorcolor=blue%
,pagecolor=blue%
,citecolor=blue%
,bookmarks=ture%
]{hyperref}

\def\qed{\hfill  \framebox(5,5){}}
\def\deg{{\rm deg}}
\def\cP{{\mathcal P}}

\expandafter\chardef\csname pre amssym.def
at\endcsname=\the\catcode`\@ \catcode`\@=11

\def\undefine#1{\let#1\undefined}
\def\newsymbol#1#2#3#4#5{\let\next@\relax
 \ifnum#2=\@ne\let\next@\msafam@\else
 \ifnum#2=\tw@\let\next@\msbfam@\fi\fi
 \mathchardef#1="#3\next@#4#5}
\def\mathhexbox@#1#2#3{\relax
 \ifmmode\mathpalette{}{\m@th\mathchar"#1#2#3}%
 \else\leavevmode\hbox{$\m@th\mathchar"#1#2#3$}\fi}
\def\hexnumber@#1{\ifcase#1 0\or 1\or 2\or 3\or 4\or 5\or 6\or 7\or 8\or
 9\or A\or B\or C\or D\or E\or F\fi}

\font\tenmsa=msam10 \font\sevenmsa=msam7 \font\fivemsa=msam5
\newfam\msafam
\textfont\msafam=\tenmsa \scriptfont\msafam=\sevenmsa
\scriptscriptfont\msafam=\fivemsa \edef\msafam@{\hexnumber@\msafam}
\mathchardef\dabar@"0\msafam@39
\def\dashrightarrow{\mathrel{\dabar@\dabar@\mathchar"0\msafam@4B}}
\def\dashleftarrow{\mathrel{\mathchar"0\msafam@4C\dabar@\dabar@}}

\font\tenmsb=msbm10 \font\sevenmsb=msbm7 \font\fivemsb=msbm5
\newfam\msbfam
\textfont\msbfam=\tenmsb \scriptfont\msbfam=\sevenmsb
\scriptscriptfont\msbfam=\fivemsb \edef\msbfam@{\hexnumber@\msbfam}
\def\Bbb#1{\fam\msbfam\relax#1}

\def\numer{{\rm numer}}

\def\deg{{\rm deg}}

\def\Content{{\rm Content}}
\def\content{{\rm Content}}

\def\pp{{\rm pp}}
\def\cP{{\mathcal P}}
\def\oh{\,{\overline h}\,}
\def\ot{\,{\overline t}\,}
\def\ox{\,{\overline x}\,}

\def\para{\vspace{6 mm}}

\def\Resultant{{\rm Res}}
\def\resultant{{\rm Res}}

\def\pp{{\rm pp}}
\def\Content{{\rm Content}}

\def\cP{{\mathcal P}}
\def\oh{\,{\overline h}\,}
\def\ot{\,{\overline t}\,}
\def\ox{\,{\overline x}\,}

 \def\numer{{\rm numer}}

\begin{document}

\begin{frontmatter}

\title{Characterization of Rational Ruled Surfaces}

\author[SPD]{Sonia P\'erez-D\'{\i}az}
\address[SPD]{Dpto. de F\'{\i}sica y Matem\'aticas,
 Universidad de Alcal\'a,
      E-28871 Madrid, Spain}
      \ead{sonia.perez@uah.es}

\author[SLY]{Li-Yong Shen\corref{cor}}
 \address[SLY]{School of Mathematical Sciences, University of
 CAS, Beijing, China}
 \ead{lyshen@ucas.ac.cn}

 \cortext[cor]{Corresponding author}

\begin{abstract}
The ruled surface is a typical modeling surface in computer aided geometric design. It is usually given in the standard parametric form. However, it can also be in the forms than the standard one. For these forms, it is necessary to determine and find the standard form. In this paper, we present algorithms to determine whether a given implicit surface is a rational ruled surface. A parametrization of the surface is computed for the affirmative case. We also consider the parametric situation. More precisely, after a given rational parametric surface is determined as a ruled one, we reparameterize it to the standard form.
\end{abstract}

\end{frontmatter}

%--------------------------------------------------------

\section{Introduction}

Parametric and implicit forms are two main representations of geometric objects. In computer aided geometric design and computer graphics, people prefer the rational parametric form for modeling design~\cite{handbook}. On the other hand, in algebraic consideration of computer algebra and algebraic geometry, people usually use the algebraic form. Since there are different advantages of parametric and implicit forms, a nature problem is to convert the forms from one to another. Converting from the implicit form to the parametric one is the parametrization problem. On the converse direction, it is the implicitization problem. There were lots of papers focused on the implicitization problem. Some typical methods were proposed using Gr\"{o}bner bases~\cite{bb85,cox98}, characteristic sets~\cite{gao92, wu90}, resultants~\cite{dixon08,sonia08} and mu-bases~\cite{buse09, chen01, dohm09}. However, there is still lack of a method having both completeness in theory and high efficiency in computation.
\para
In general, the parametrization problem is more difficult than the implicitization problem.  Only some of the algebraic curves and surfaces have rational parametric representations. For the curves, people have proposed different methods such as parametrization based on resolvents~\cite{gao92b}, by lines or adjoint curves~\cite{libro} (see Chapter 4) and using  canonical divisor~\cite{van97}.
For a general surface, it is still an opening problem to propose an efficient parametrization algorithm. But to meet the practical demands, people had to design the parametrization algorithms for some commonly used surfaces. Sederberg and Snively~\cite{seds87} proposed four parametrization methods for cubic algebraic surfaces. One of them was based on finding two skew lines that lie in the surface. Sederberg~\cite{sed90b} and Bajaj et al.~\cite{bajaj98} expanded this method.
In~\cite{handbook02}, a method to parameterize a quadric was given using a stereographic projection. Berry et al.~\cite{berry01} unified the implicitization and parametrization of a nonsingular cubic surface with Hilbert-Burch matrices. These methods were designed for some special surfaces. In~\cite{schicho97}, Schicho gave deeper analysis in parametrization problem. He gave more contributions on theoretical analysis than practicable computation. Therefore, it is still necessary to find the efficient parametrization algorithm for certain commonly used surfaces.

\para

The ruled surface is an important surface widely used in computer aided geometric design and geometric modeling (see \cite{ARTV11, buse09,chen03, chen01,chen11,dohm09,izu03, liu06, li08,sonia08, shen10, shen12}). Using the $\mu$-bases method, Chen et al.~\cite{chen01} gave an implicitization algorithm for the rational ruled surface. The univariate resultant was also used to compute the implicit equations efficiently~\cite{sonia08, shen10}.
For a given rational ruled surface, people could find a simplified reparametrization which did not contain any non-generic base point and had a pair of directrices with the lowest possible degree~\cite{chen03}. Bus\'{e} and Dohm took more look at the ruled surface using $\mu$-bases~\cite{buse09,dohm09} respectively.
Li et al.~\cite{li08} could find a proper reparametrization of an improper parametric ruled surface.
Andradas et al. presented an algorithm to decide whether a proper rational parametrization of a ruled surface could be properly reparametrized over a real field~\cite{ARTV11}.
The ruled surfaces had been used for geometric modeling of architectural freeform design in~\cite{liu06}. The collision and intersection of the ruled surfaces were discussed in~\cite{chen11, shen12}. And S. Izumiya~\cite{izu03} studied  the cylindrical helices and Bertrand curves as curves on ruled surfaces.
In these papers, the ruled surface was given in standard parametric form ${\cal Q}(t_1, t_2)={\cal M}(t_1)+t_2{\cal N}(t_1)\in {\Bbb K}(t_1, t_2)^3$. It means the rational ruled surface was preassigned in the discussions. But in general modeling design, such as data fitting, the type of approximate surface may be not known. Then a problem is, for a given parametric surface not being standard form of the ruled surface, how to determine whether it is a ruled surface. If the answer is affirmative, the successive problem is then to find a standard parametric form. In this paper, we would like to consider the determination and reparametrization of the parametric ruled surface.

\para

Go back to parametrization, the implicit surfaces are often introduced in algebraic analysis. And they can also come directly from modeling design since they have more geometric features and topologies than those of the parametric surfaces (see \cite{handbook, turk02}). As we know, there was no paper discussing the parametrization of an implicit rational ruled surface. Here, we would like to consider the parametrization problem of the implicit rational ruled surface. Precisely, for a given algebraic surface, we first determine whether it is a rational ruled surface, and in the affirmative case, we compute a rational parametrization in standard form.
Our discussion is benefited from the standard presentation of the rational ruled surface. Since the parameter $t_2$ is linear, we can construct a birational parameter transformation to simply the given parametric ruled surface. By the linearity again, $t_2$ is always solvable such that we can project the surface to the rational parametric curve. And these two main techniques help us to give the determination and (re)parametrization algorithms. The main theorems are all proved constructively, and the algorithms are then presented naturally.

\para

The paper is organized as follows. First, some necessary preliminaries of ruled surfaces are presented in Section~2. In Section~3, we determine whether a given implicit surface is a rational ruled surface, and in the affirmative case, we compute a rational parametrization in standard form for it. In Section~4, we focus on the parametric surface including determination and reparametrization. Finally, we conclude with
Section~5, where we propose topics for further study.

\section{Preliminaries on Ruled Surfaces}

Let $\cal V$ be a ruled surface over an algebraically closed field of characteristic zero $\Bbb K$, and let $f(\ox)\in {\Bbb K}[\ox],\,\,\ox=(x_1,x_2,x_3)$ be the polynomial defining implicitly the surface $\cal V$.
%We assume that $\cal V$ is not the plane $x_i-c=0,\,c\in {\Bbb C}$ for   $i=1,2,3$, and $\cal V$ is not a cylinder  over any of the
%coordinate planes of ${\Bbb K}^3$.  That is, $\deg_{x_i}(f)>0$, for $i=1,2,3$. If $\deg_{x_3}(f)=0$ (similarly if  $\deg_{x_1}(f)=0$  or $\deg_{x_2}(f)=0$), we may compute a proper %parametrization  $(p(t_1), q(t_1))$  of the curve defined by the polynomial $f(x_1, x_2)=0$. Then, $\cP(t_1, t_2)=(p(t_1), q(t_1), t_2)\in {\Bbb K}(\ot)^3$ is a proper parametrization %of $\cal V$.

\para

%Under these conditions, we get that a proper parametrization of a ruled surface $\cal V$, that is not a cylinder,  is given by
\noindent
A proper parametrization of a rational ruled surface $\cal V$  is given by
\begin{equation}\label{eq-standardform}{\cal Q}(\ot)=(m_1(t_1)+t_2n_1(t_1), m_2(t_1)+t_2n_2(t_1), m_3(t_1)+t_2n_3(t_1))\in {\Bbb K}(\ot)^3,\quad \ot=(t_1, t_2)\end{equation}
where there exists at least one  $i\in \{1,2,3\}$ such that $n_i\not=0$ (if $n_j=0$ for $j=1,2,3$, then $\cal V$ is a space curve). We refer to this parametrization, as the {\sf standard form parametrization of $\cal V$.}
%where  at least $n_in_j\not=0$, for $i\not=j$ and $i,j\in \{1,2,3\}$. We note that if we do not assume that  $\cal V$ is not a plane $x_i-c=0,\,c\in {\Bbb C}$ for   $i=1,2,3$, and $\cal V$ is not a cylinder  over any of the
%coordinate planes, we only can ensure that there exists $i \in \{1,2,3\}$ such that $n_i\not=0$ (if $n_1=n_2=n_3=0$, then $\cal Q$ parametrizes a space curve).

\para

\noindent
Note that if $n_3\not=0$, the surface $\cal V$ admits a proper parametrization of the form
\begin{equation}\label{eq-standardreducedform}{\cal P}^3(\ot)=(p_{13}(t_1)+t_2q_{13}(t_1), p_{23}(t_1)+t_2q_{23}(t_1), t_2)\in {\Bbb K}(\ot)^3,\end{equation}
where $q_{k3}=n_k/n_3\not=0$, for some $k=1,2$. Such a parametrization is obtained by performing the birational transformation
\[(t_1, t_2)\rightarrow \left(t_1, \frac{t_2-m_3(t_1)}{n_3(t_1)}\right).\]
One may reason similarly, if $n_1\not=0$ or $n_2\not=0.$
In the following, we refer to parametrization ${\cal P}^i$ as the  {\sf standard  reduced form parametrization of $\cal V$}.

\para

\noindent
Under these conditions, we distinguish two different cases:
\begin{itemize}
\item If $n_1n_2n_3\not=0$, then $q_{13}q_{23}\not=0$, and
\[{\cal P}^3(t_1, 0)=(p_{13}, p_{23},0)
,\quad {\cal P}^3\left(t_1, -\frac{p_{13}}{q_{13}}\right)=\left(0, p_{23}-\frac{p_{13}}{q_{13}}q_{23}, -\frac{p_{13}}{q_{13}}\right),\quad\]\[ {\cal P}^3\left(t_1, -\frac{p_{23}}{q_{23}}\right)=\left(p_{13}-\frac{p_{23}}{q_{23}}q_{13}, 0,-\frac{p_{23}}{q_{23}}\right).\]
 parametrize a rational plane  curve defined by a factor of the polynomials
$$f_0^{12}(x_1, x_2)=f(x_1,x_2,0),\quad f_0^{23}(x_2, x_3)=f(0,x_2,x_3), \quad f_0^{13}(x_1, x_3)=f(x_1,0,x_3),$$
respectively. We denote by ${\cal C}^{ij}$ these rational curves, where $ij\in \{12,23,13\}$.
\item Let us assume that $n_2=0$, and $n_1n_3\not=0$. Then,
\[{\cal Q}(\ot)=(m_1(t_1)+t_2n_1(t_1), m_2(t_1), m_3(t_1)+t_2n_3(t_1))\in {\Bbb K}(\ot)^3.\]
Since $n_3\not=0,$
\[{\cal P}^3(\ot)=(p_{13}(t_1)+t_2q_{13}(t_1),m_2(t_1), t_2),\quad q_{13}=n_1/n_3\not=0,\]
and  ${\cal P}^3(t_1, 0)$, $ {\cal P}^3\left(t_1, -\frac{p_{13}}{q_{13}}\right)$ parametrize a  rational plane curve defined by a factor of the polynomials
$$f_0^{12}(x_1, x_2)=f(x_1,x_2,0),\quad f_0^{23}(x_2, x_3)=f(0,x_2,x_3)$$
respectively.  We denote by ${\cal C}^{ij}$ these rational curves, where $ij\in \{12,23\}$.
\end{itemize}

%----------------------------------------------------------------

\section{Implicitly Rational Ruled Surfaces}

In this section, we consider a surface  $ \cal V$ defined over an algebraically closed field of characteristic zero $\Bbb K$, and a polynomial $f(\ox)\in {\Bbb K}[\ox]$ defining implicitly  $ \cal V$.  \\

In the following, we analyze whether $\cal V$ is a rational ruled surface. In the affirmative case, we compute a rational proper parametrization of $\cal V$ in the standard form given by the equation~\eqref{eq-standardform}.
%that is,
%\[{\cal Q}(\ot)=(m_1(t_1)+t_2n_1(t_1), m_2(t_1)+t_2n_2(t_1), m_3(t_1)+t_2n_3(t_1))\in {\Bbb K}(\ot)^3.\]
%
For this purpose,   we   denote by $\numer(R)$, the numerator of a rational function $R\in {\Bbb K}(x_1,x_2,\ldots,x_n)$.

\para

\begin{theorem}\label{T-implicharac2} A surface $\cal V$ defined implicitly by a polynomial  $f(\ox)\in {\Bbb K}[\ox]$  is a rational ruled surface if and only if the following statements holds:
\begin{itemize}
\item[1.]  There exists a rational plane curve    defined by a factor of the polynomial $f_0^{ij}$,  for some $ij\in \{12,13,23\}$ (see Section 2). Let $ij=12$,  and let $\cP^{12}=(p_1,p_2)\in {\Bbb K}(t_1)^2$ be a rational proper parametrization of the rational plane curve ${\cal C}^{12}$ defined  by the factor of $f_0^{12}$.
\item[2.] There exist  $(q_1,q_2)\in {\Bbb K}(t_1)^2$ and $S\in {\Bbb K}(t_1)\setminus{\Bbb K}$ such that $g(q_1(t_1),q_2(t_1), S(t_1), t_2)=0$, where  $g(x_1,x_2,x_3, t_2)=\numer(f(p_1(x_3)+t_2x_1,p_2(x_3)+t_2x_2,t_2))$.
  In this case, \[{\cal P}(\ot)=(p_1(S(t_1))+t_2q_1(t_1), p_2(S(t_1))+t_2q_2(t_1),t_2)\] is a rational proper parametrization of $\cal V$.
 \end{itemize}
\end{theorem}

\vspace*{2mm}

\noindent {\bf Proof.}  It is clear that if statements $1$ and $2$ hold, then  $\cal V$ is a rational ruled surface. Reciprocally, let $\cal V$ be a rational ruled surface. Then a proper parametrization of $\cal V$ is given by  the standard form parametrization~\eqref{eq-standardform}. That is,
\[{\cal Q}(\ot)=(m_1(t_1)+t_2n_1(t_1), m_2(t_1)+t_2n_2(t_1), m_3(t_1)+t_2n_3(t_1))\in {\Bbb K}(\ot)^3.\] We  assume  that $n_3\not=0$ (see Section 2 and  Remark~\ref{R-ij2}).  Thus,
 ${\cal Q}(t_1, -m_3/n_3)$ parametrizes ${\cal C}^{12}$, and statement 1 holds.  Let us prove that statement 2 holds. For this purpose, we consider $\cP^{12}=(p_1,p_2)\in {\Bbb K}(t_1)^2$ a rational proper  parametrization of ${\cal C}^{12}$. In addition, since $n_3\not=0$, we have that  $\cal V$ admits a  standard reduced  form  parametrization given in the equation~\eqref{eq-standardreducedform}. That is,
 \[{\cal P}^3(\ot)=(p_{13}(t_1)+t_2q_{13}(t_1), p_{23}(t_1)+t_2q_{23}(t_1), t_2)\in {\Bbb K}(\ot)^3.\] Observe that ${\cal P}^3(t_1, 0)=(p_{13},p_{23})\in {\Bbb K}(t_1)^2$ is a rational  parametrization of ${\cal C}^{12}$. Then, since $\cP^{12}$ is a proper parametrization of ${\cal C}^{12}$, there exists $S\in {\Bbb K}(t_1)\setminus{\Bbb K}$ such that    $\cP^{12}(S)=(p_1(S),p_2(S))=(p_{13},p_{23})$.   Thus, since ${\cal P}^3$ parametrizes properly $\cal V$, we  have that
\[{\cal P}(\ot)=(p_1(S(t_1))+t_2q_1(t_1), p_2(S(t_1))+t_2q_2(t_1), t_2)\in {\Bbb K}(\ot)^3,\,\, \mbox{where} \quad (q_1, q_2)=(q_{13}, q_{23})\]
is  a proper parametrization of $\cal V$, and    $(q_1,q_2)\in {\Bbb K}(t_1)^2$, $S\in {\Bbb K}(t_1)\setminus{\Bbb K}$ satisfy  that  $g(q_1(t_1),q_2(t_1), S(t_1), t_2)=\numer(f(p_1(S(t_1))+t_2q_1(t_1),p_2(S(t_1))+t_2q_2(t_1),t_2))=0$.
\qed

\para

\begin{remark}\label{R-ij2} Note that in  the Theorem~\ref{T-implicharac2}, we assume that ${\cal C}^{12}$ is the rational plane curve satisfying statement~1. In addition, in the proof of theorem, when a rational ruled surface is given, we consider a parametrization $\cal Q$ in standard form~\eqref{eq-standardform} such that  $n_3\not=0$ (from this assumption, we have that  ${\cal Q}(t_1, -m_3/n_3)$ parametrizes ${\cal C}^{12}$).\\
Theorem~\ref{T-implicharac2} is proved similarly if a different rational  plane  curve ${\cal C}^{ij}$ is considered in statement~1, and a different polynomial  $n_i$ satisfies that  $n_i\not=0$ (see Section 2). In addition, if  $n_1\not=0$, we get that $g(\ox, t_2)=\numer(f(t_2,p_1(x_3)+t_2x_1,p_2(x_3)+t_2x_2))$, and if  $n_2\not=0$, we get that $g(\ox, t_2)=\numer(f(p_1(x_3)+t_2x_1,t_2,p_2(x_3)+t_2x_2))$.\\

\noindent
In the following, we assume that $n_3\not=0$, and then we are in the conditions of Theorem~\ref{T-implicharac2}.  This requirement can always be achieved by applying a linear
transformation to $\cal V$,  and therefore it is not a loss of
generality for our purposes since one can always undo the linear
transformation once  $\cal P$ has been computed.
\end{remark}

\para

In Corollary~\ref{C-implicharac2}, we prove that statement~2 in Theorem~\ref{T-implicharac2} is equivalent to check the rationality of a space curve, and to compute, in the affirmative case, a rational parametrization.

\para

\begin{corollary}\label{C-implicharac2} Let $\cal V$ be a surface defined implicitly by a polynomial  $f(\ox)\in {\Bbb K}[\ox]$ and such that statement~1 in Theorem~\ref{T-implicharac2} holds. $\cal V$ is a rational ruled surface if and only if  the coefficients of the polynomial  $g(\ox, t_2)$ w.r.t the variable $t_2$ define a  rational space curve $\cal D$. In this case, ${\cal Q}(t_1)=(q_1(t_1),q_2(t_1), S(t_1))\in {\Bbb K}(t_1)^3$, where $S\not\in {\Bbb K}$, is a rational parametrization of $\cal D$.
 \end{corollary}

\vspace*{2mm}

\noindent {\bf Proof.} First, we write
\[g(\ox, t_2)=h_0(\ox)+h_1(\ox)t_2+\cdots+h_n(\ox)t_2^n,\]
and we prove that $h_0=0$, the only factor in ${\Bbb K}[t_2]$ dividing $g$ is $t_2^r$ for some $r\in {\Bbb N}$, and there exist at least two polynomials, $h_i, h_j$, such that $h_i\not=h_j$. Indeed: \begin{enumerate}
\item[a.]  $h_0=0$: since $g(\ox, t_2)=\numer(f(p_1(x_3)+t_2x_1,p_2(x_3)+t_2x_2,t_2))$, and $f(p_1(x_3),p_2(x_3),0)=0$, we deduce that $t_2$ divides $g$.
\item[b.]  The only factor in ${\Bbb K}[t_2]$ dividing $g$ is $t_2^r,\,r\in {\Bbb N}$: let $c\in {\Bbb K}\setminus\{0\}$ be such that $g(\ox, c)=0$. Since $g(\ox, t_2)=\numer(f(p_1(x_3)+t_2x_1,p_2(x_3)+t_2x_2,t_2))$, we deduce that $f(p_1(x_3)+cx_1,p_2(x_3)+cx_2,c)=0$, for every $x_1, x_2$. Then, $\cal V$ is
the plane defined by the equation $x_3-c=0$ which is impossible because we have assumed that $n_3\not=0$.
\item[c.] There  exist at least two polynomials, $h_i, h_j$, such that $h_i\not=h_j$. Let us assume that this statement does not hold. From statement $b$, this implies that $g(\ox, t_2)=t_2^r h(\ox).$ In addition, reasoning as in statement $b,$ we have that $h\not=0$. From the equality $$g(\ox, t_2)=\numer(f(p_1(x_3)+t_2x_1,p_2(x_3)+t_2x_2,t_2))=t_2^r h(\ox),$$ and deriving w.r.t. $x_1, x_2, x_3$,  we get that, up to factors in ${\Bbb K}[x_3]\setminus{\{0\}}$,
    \[f_{x_1}(p_1(x_3)+t_2x_1,p_2(x_3)+t_2x_2,t_2)t_2=t_2^r h_{x_1}(\ox),\]\[ f_{x_2}(p_1(x_3)+t_2x_1,p_2(x_3)+t_2x_2,t_2)t_2=t_2^r h_{x_2}(\ox),\]\[ f_{x_1}(p_1(x_3)+t_2x_1,p_2(x_3)+t_2x_2,t_2)p'_1+f_{x_2}(p_1(x_3)+t_2x_1,p_2(x_3)+t_2x_2,t_2)p'_2=t_2^r h_{x_3}(\ox),\]
     where $p_{var}$ represents the partial derivative of a polynomial $p$ w.r.t the variable $var$.   Thus, up to factors in ${\Bbb K}[x_3]\setminus{\{0\}}$,
     \[ h_{x_1}(\ox)p'_1(x_3)+h_{x_2}(\ox)p'_2(x_3)=t_2 h_{x_3}(\ox)\]
     which implies that $h_{x_3}=0$, and then $h\in {\Bbb K}[x_1,x_2]$. Hence, $$g(\ox, t_2)=\numer(f(p_1(x_3)+t_2x_1,p_2(x_3)+t_2x_2,t_2))=t_2^r h(x_1, x_2).$$
     Let $\eta_i=(a_i,b_i)\in {\Bbb K}^2$ be such that $h(\eta_i)=0,\,i=1,2$, and $\eta_1\not=\eta_2$. Then, $g(a_i, b_i, x_3, t_2)=\numer(f(p_1(x_3)+t_2a_i,p_2(x_3)+t_2b_i,t_2))=0$, and thus  ${\cal Q}_i=(p_1(x_3)+t_2a_i,p_2(x_3)+t_2b_i,t_2)$ parametrizes $\cal V$ which implies that
     ${\cal Q}_1(U,V)={\cal Q}_2$, where $(U, V)\in ({\Bbb K}(t_1)\setminus{\Bbb K})^2$. Thus, $V=t_2$ and $\eta_1=\eta_2$ which is impossible. Therefore, there are not two different points on the curve defined by $h$. Hence, $h\in {\Bbb K}\setminus\{0\}$ and then, up to constants in ${\Bbb K}\setminus\{0\}$, $$g(\ox, t_2)=\numer(f(p_1(x_3)+t_2x_1,p_2(x_3)+t_2x_2,t_2))=t_2^r.$$ This is impossible, because if we consider $\eta:=(a_1, a_2, a_3)\in {\Bbb K}^3,\,a_3\not=0$,  with $f(\eta)=0$ (observe that this point exists because  $n_3\not=0$, and then $\cal V$ is not the plane $x_3=0$), we have that  $f(p_1(x_3)+a_3x^0_1,p_2(x_3)+a_3x^0_2,a_3)=0$, where $x^0_j=(a_j-p_j(x_3))/a_3,\,j=1,2$. Thus, $g(x^0_1,x^0_2,x_3,a_3)=a_3^r=0$ which is a contradiction.
     %which implies that $g\not\in{\Bbb K}[y]$. Under these conditions, we consider $(\ell_1,\ell_2,\ell_3)\in {\Bbb K}^3$ such that $h(\ell_1,\ell_2,\ell_3)=0$,  and $(p_1,p_2)$ is defined at $\ell_3$. Since $g\not\in{\Bbb K}[y]$, we have that this point exists. Then,   $f(p_1(\ell_3)+t_2\ell_1,p_2(\ell_3)+t_2\ell_2,t_2)=0$ which is impossible  because  $\cal V$ is not an space curve.
    \end{enumerate}
Taking into account these results, we prove the corollary. First,  if $\cal V$ is a rational ruled surface, statement~2 in Theorem~\ref{T-implicharac1} holds, and then $g(q_1(t_1),q_2(t_1), S(t_1), t_2)=0$.
Since ${\cal Q}(t_1):=(q_1,q_2, S)\in {\Bbb K}(t_1)^3$ does not depend on $t_2$, we get that $h_i({\cal Q})=0,\,i\in \{0,\ldots, n\}$ which implies that $\cal Q$ is a rational parametrization of the space  curve $\cal D$  defined by the polynomials $h_i,\,i\in \{0,\ldots, n\}$ (see statements a, b, c above).\\
Reciprocally, let $(U(t_1),V(t_1), W(t_1))\in {\Bbb K}(t_1)^3$ be a rational parametrization of $\cal D$. Then, $h_i(U,V, W)=0,\,i\in \{0,\ldots, n\}$ which implies that  $g(U(t_1),V(t_1), W(t_1), t_2)=0$. Hence, $f({\cal P})=0,$ where
\[{\cal P}(\ot)=\left(p_1(W(t_1))+t_2U(t_1), p_2(W(t_1))+t_2V(t_1),t_2\right).\]
%We observe that $W\not\in {\Bbb K}$. Indeed, let us assume that  $W=a \in {\Bbb K}$. Since $g(U(t_1),V(t_1), a, t_2)=0$, we get that
 \qed

\begin{remark}\label{R-g} From the proof in Corollary~\ref{C-implicharac2}, we have that \[g(\ox, t_2)=t_2^r(h_1(\ox)+\cdots+h_{m+1}(\ox)t_2^{m+1}),\quad r,m\in {\Bbb N},\]
 and there exist at least two polynomials, $h_i, h_j$, such that $h_i\not=h_j$. Under these conditions, the space curve ${\cal D}$ is defined at least by the polynomials $h_i,$ and $h_j$. Taking into account that any space curve can be birationally projected onto a
plane curve, one may apply results in  Chapter 6 in \cite{libro} to compute  ${\cal Q}(t_1)=(q_1(t_1),q_2(t_1), S(t_1))$.
 % In order to compute  ${\cal Q}(t_1)=(q_1(t_1),q_2(t_1), S(t_1))$, one may proceed as follows:
%\begin{enumerate}
%\item If $m=1$, since $h_1\not=h_2$,  we  determine $R(x_1, x_2)=\resultant_y(\bar{h}_1, \bar{h}_2)$, where $\bar{h}_i=h_i/\gcd(h_1,h_2)$,\,$i=1,2$. Then, we compute a parametrization, $(q_1, q_2)$, of the rational curve defined  by a factor  of $R$ such that there exists $S$ satisfying that $\gcd(\bar{h}_1(q_1,q_2, y), \bar{h}_2(q_1,q_2, y))(S)=0$  (note that $R(q_1,q_2)=0$).   For further details  see e.g.  Chapter 6 in \cite{libro}.
%\item If  $m>1$, one may proceed as above with two polynomials and checking at the end that $g({\cal Q},t_2)=0$. Additionally, one also may  solve the equations $h_i=0$  on the variables $x_1, x_2$ and $y$, and look for a rational solution  (note that $h_i(q_1,q_2, S)=0$).
%\end{enumerate}
\end{remark}

\para

Taking into account Theorem~\ref{T-implicharac2} and  Corollary~\ref{C-implicharac2}, one may check whether $\cal V$ is a rational ruled surface, and in the affirmative case to compute a rational proper parametrization in standard reduced form (see equation given in~\eqref{eq-standardreducedform}).
\para

\noindent
\fbox{{\sf Algorithm Parametrization of a Rational Ruled Surface 1.}}
\begin{itemize}
\item {\bf Input:} A surface $\cal V$ defined by an irreducible polynomial $f(\ox)\in {\Bbb K}[\ox]$.
\item {\bf Output:} the message ``{\tt $\cal V$ is not a rational ruled surface}" or
a proper parametrization ${\cal P}$ of ``{\tt the  rational  ruled surface $\cal V$}".
\end{itemize}
 \begin{itemize}
 \item[1.] Check whether any component of the curve  defined by the polynomial $f_0^{ij}$,  for $ij\in \{12,13,23\}$, is rational. In the affirmative case, we assume that $ij = 12$  (see Remark~\ref{R-ij2}). Otherwise,   {\sc Return} ``{\tt ${\cal V}$ is not a  rational  ruled surface}".
\item[2.]  Compute ${\cal P}^{12}=(p_1,p_2)\in {\Bbb K}(t_1)^2$ a rational proper parametrization of ${\cal C}^{12}$. For this purpose, apply for instance the results in Sections 4.7 and 4.8 in \cite{libro}.
\item[3.]  Let  $g(\ox, t_2)=\numer(f(p_1(x_3)+t_2x_1,p_2(x_3)+t_2x_2,t_2))$.  Check whether the coefficients of the polynomial  $g(\ox, t_2)$ w.r.t the variable $t_2$  define a rational space curve $\cal D$. In the affirmative case,  compute ${\cal Q}=(q_1,q_2, S)\in {\Bbb K}(t_1)^3$  a rational parametrization of $\cal D$ (see Remark~\ref{R-g}), and {\sc Return}
    \[{\cal P}=(p_1(S(t_1))+t_2q_1(t_1),p_2(S(t_1))+t_2q_2(t_1),t_2)\in {\Bbb K}(\ot)^3\]
   ``{\tt  is a proper parametrization of the  rational  ruled surface  ${\cal V}$ in standard reduced form}". Otherwise, go to Step 1, and consider a different rational component and apply again the algorithm. If there have no more rational components, {\sc Return} ``{\tt $\cal V$ is not a  rational  ruled surface}".
\end{itemize}

\para

\begin{remark}\label{R-Alg2} If
${\cal P}^{12}$ and $\cal Q$ have coefficients in a field $\Bbb L$, then the output parametrization $\cal P$  also has coefficients in  $\Bbb L$. Then, in particular, if we compute the proper parametrizations, ${\cal P}^{12}$ and $\cal Q$,  in the smallest possible field extension of the ground field (see Chapter 5 in \cite{libro}), the output parametrization $\cP$   belongs to this
smallest possible field extension. For practical applications, we may  consider a surface over the real field $\Bbb R$ and to compute $\cP$ with real coefficients, if it is possible. For this purpose, we may apply the results  in \cite{libro} (see Chapter 7), and compute ${\cal P}^{12}$ and $\cal Q$  over the reals (if it is possible). In this sense, we observe that we return a proper parametrization in standard reduced form over $\Bbb R$ (if it exists). Compare with the results in \cite{ARTV11}, where  an algorithm to decide whether a proper rational parametrization of a ruled surface can be properly reparametrized over $\Bbb R$ is presented. The output in this paper is not necessarily given in standard form.\\

\noindent
Finally we observe that if ${\cal P}^{12}$ and $\cal Q$  are polynomial (see Section 6.2 in \cite{libro}), then   $\cP$ is also polynomial.
 \end{remark}

 \para

 In the following, we illustrate {\sf Algorithm Parametrization of a Rational Ruled Surface 1}, with two examples.

\para

\begin{example} Consider the surface $\cal V$ over the complex field $\Bbb C$ defined implicitly by the polynomial $$f(\ox)=x_1^2+x_2^2+x_3^2-1.$$ Let us apply {\sf Algorithm Parametrization of a Rational Ruled Surface 1}. For this purpose, we first observe that $f(x_1,x_2,0)=x_1^2+x_2^2-1$ is a rational plane curve, and we compute a rational proper parametrization of this curve
\[{\cal P}^{12}(t_1)=(p_1(t_1),p_2(t_1))=\left(\frac{2t_1}{t_1^2+1}, \frac{t_1^2-1}{t_1^2+1}\right)\in {\Bbb R}(t_1)^2.\]
Now, we determine the polynomial  $g(\ox, t_2)=\numer(f(p_1(x_3)+t_2x_1,p_2(x_3)+t_2x_2,t_2))$. We get
$$g(\ox, t_2)=t_2(2x_3^2x_2+t_2x_2^2x_3^2+t_2x_3^2+t_2x_1^2x_3^2+4x_3x_1+t_2+t_2x_1^2+t_2x_2^2-2x_2).$$
The  coefficients of the polynomial $g(\ox, t_2)$ w.r.t the variable $t_2$, are
\[ h_1(\ox)= 2 x_2 x_3^2  + 4 x_3 x_1 - 2 x_2,\quad
              h_2(\ox)= x_1^2  x_3^2  + x_2^2  x_3^2  + x_3^2  + x_1^2  + x_2^2  + 1.\]
Note that $\gcd(h_1, h_2)=1$. These polynomials define implicitly the rational space curve ${\cal D}$.  We apply Remark~\ref{R-g},  and   we compute a rational parametrization of  $\cal D$:
%$$R(x_1, x_2)=\resultant_y(h_1, h_2)=16(x_1^2+x_2^2+1)^2(x_1^2+x_2^2).$$
%We parametrize the rational curve defined by the equation $x_1^2+x_2^2+1=0$ (the other factor, $x_1^2+x_2^2$, does not provide $S$ such that $\gcd(h_1(q_1,q_2, y), h_2(q_1,q_2, y))(S)=0$). We obtain $$(q_1, q_2)=\left(\frac{I(1+t_1^2)}{2t_1}, \frac{-1+t_1^2}{2t_1}\right)\in {\Bbb C}(\ot)^2.$$ Now, we compute $S$ such that $\gcd(h_1(q_1,q_2, y), h_2(q_1,q_2, y))(S)=0$. We get that $S=-\frac{I(-1+t_1)}{1+t_1},$ and then
$${\cal Q}(t_1)=(q_1(t_1),q_2(t_1), S(t_1))=\left(\frac{I(1+t_1^2)}{2t_1}, \frac{-1+t_1^2}{2t_1},-\frac{I(-1+t_1)}{1+t_1}\right)\in {\Bbb C}(t_1)^3.$$
Thus, ${\cal P}=(p_1(S(t_1))+t_2q_1(t_2),p_2(S(t_1))+t_2q_2(t_2),t_2)=$
\[\left(\frac{I(1-t_1^2+t_2+t_2t_1^2)}{2t_1}, \frac{-1-t_1^2-t_2+t_2t_1^2}{2t_1}, t_2\right)\in {\Bbb C}(\ot)^3\]
   ``{\tt  is a proper parametrization of the  rational  ruled surface  ${\cal V}$  in stan\-dard\- reduced form}". Taking into account Remark~\ref{R-Alg2}, we conclude that there not exist a parametrization over $\Bbb R$ in standard reduced form of the  rational  ruled surface $\cal V$. However, one may apply results in \cite{ARTV11} to decide whether $\cal V$ can be parametrized over the reals and to compute, in the affirmative case, a real parametrization.
\end{example}

\para

\begin{example} Let $\cal V$ be the surface  defined implicitly by the polynomial \\

\noindent
$f(\ox)=-49x_2x_1^3-799x_3x_2x_1^2+20x_2x_1^2+2x_2^2x_1^2+980x_3x_1^2-2205x_3^2x_1^2+x_2^3x_1-33750x_3^3x_1-400x_3x_1+606x_3x_2x_1-5x_2^2x_1-68x_3x_2^2x_1-1747x_3^2x_2x_1-25x_3^2x_1+x_2^3x_3-25x_2^2x_3^2+1396x_2x_3^2-1120x_3^2-48915x_3^4-5190x_3^3-4237x_2x_3^3-14x_2^2x_3\in {\Bbb C}[\ox].$\\

\noindent
Let us apply {\sf Algorithm Parametrization of a Rational Ruled Surface 1}. For this purpose, we first observe that $f_0^{12}(x_1,x_2)=-x_1x_2(49x_1^2-20x_1-2x_1x_2-x_2^2+5x_2)$. The curve defined by the equation $49x_1^2-20x_1-2x_1x_2-x_2^2+5x_2=0$ is rational. We compute a rational proper parametrization of this curve. We get,
\[{\cal P}^{12}(t_1)=(p_1(t_1),p_2(t_1))=\left(\frac{-\sqrt{2}t_1(-5+t_1)}{5(4t_1-10+5\sqrt{2})}, \frac{(50+5\sqrt{2})t_1(-20+100\sqrt{2}+49t)}{1225(4t_1-10+5\sqrt{2})}\right).\]
Now, we determine the polynomial $g(\ox, t_2)=\numer(f(p_1(x_3)+t_2x_1,p_2(x_3)+t_2x_2,t_2))$, and we consider the space curve $\cal D$ defined by the  coefficients of  $g$ w.r.t $t_2$. By applying  Remark~\ref{R-g}, we have that $\cal D$ is rational, and we compute a rational parametrization of  $\cal D$:
%and we solve the equations defined by the coefficients of  $g$ w.r.t $t_2$ (we have more than two polynomials defined by the coefficients of $g$; see Remark~\ref{R-g}). We obtain the following rational solution
$${\cal Q}(t_1)=\left(q_1(t_1),q_2(t_1), S(t_1)\right)=\left(\frac{25-6t_1-25\sqrt{2}+5\sqrt{2}t_1}{t_1}, \frac{-9 t_1(1+\sqrt{2})}{-5+t_1}, t_1\right).$$
Thus,
  \[{\cal P}(\ot)=(p_1(t_1)+t_2q_1(t_2),p_2(t_1)+t_2q_2(t_2),t_2)=\]\[\left(\frac{-5  \sqrt{2} t_1^2+ \sqrt{2} t_1^3-1050 t_2 t_1+120 t_2 t_1^2+900 t_2 t_1  \sqrt{2}-100 t_2 t_1^2  \sqrt{2}+2500 t_2-1875 t_2  \sqrt{2}}{5t_1 (-4 t_1+10-5  \sqrt{2})},\right.\]\[\left. \frac{t_1 (15  \sqrt{2} t_1-100  \sqrt{2}-50 t_1+10 t_1^2+ \sqrt{2} t_1^2-180 t_2 t_1-180 t_2 t_1  \sqrt{2}+225 t_2  \sqrt{2})}{5(-5+t_1) (4 t_1-10+5  \sqrt{2})}, t_2\right)\]
   ``{\tt  is a proper  parametrization of the  rational  ruled surface  ${\cal V}$ over $\Bbb R$  in standard reduced form}".
      \end{example}

\para

 In the following, we present a new result that characterizes  rational  ruled surfaces defined implicitly. From this theorem, we obtain a new algorithm that allows to analyze whether  $\cal V$ is a  rational   ruled surface and in the affirmative case, to compute a rational parametrization of $\cal V$. In this new approach, instead to decide whether the polynomial $g$ defines a rational space curve $\cal D$, we only need to decide whether a new polynomial, constructed directly from two rational parametrizations of two plane curves, defines a rational plane curve. That is, we do not need to work on the space. This new approach plays an important role to deal with surfaces defined parametrically in Section 4.\\

  In the following new approach, we need to assume that $\cal V$ is not the plane $x_i-c=0,\,c\in {\Bbb C}$ for   $i=1,2,3$, and $\cal V$ is not a cylinder  over any of the
coordinate planes of ${\Bbb K}^3$.  That is, $\deg_{x_i}(f)>0$, for $i=1,2,3$. If $\deg_{x_3}(f)=0$ (similarly if  $\deg_{x_1}(f)=0$  or $\deg_{x_2}(f)=0$), we may compute a proper parametrization  $(p(t_1), q(t_1))$  of the plane curve defined by the polynomial $f(x_1, x_2)=0$. Then, $\cP(\ot)=(p(t_1), q(t_1), t_2)\in {\Bbb K}(\ot)^3$ is a proper parametrization of $\cal V$.

\para

 Under these conditions, and taking into account Section 2, we get that a proper parametrization of a  rational  ruled surface $\cal V$, that is not a cylinder neither a plane,  is given by the standard form parametrization given in the equation~\eqref{eq-standardform},
%\[{\cal Q}(\ot)=(m_1(t_1)+t_2n_1(t_1), m_2(t_1)+t_2n_2(t_1), m_3(t_1)+t_2n_3(t_1))\in {\Bbb K}(\ot)^3,\]
where  at least there exist $i,j\in \{1,2,3\}$,\,$i\not=j,$ such that $n_in_j\not=0$. We note that if we do not assume that  $\cal V$ is not a plane $x_i-c=0,\,c\in {\Bbb C}$ for   $i=1,2,3$, and $\cal V$ is not a cylinder  over any of the
coordinate planes, we only can ensure that there exists $i \in \{1,2,3\}$ such that $n_i\not=0$ (if $n_1=n_2=n_3=0$, then $\cal Q$ parametrizes a space curve).

\para

In Theorem~\ref{T-implicharac1}, we characterize whether a surface defined implicitly is a  rational  ruled surface by analyzing the rationality of two plane curves defined directly from the input surface. In addition, from the parametrization of these plane curves, we can compute a rational proper parametrization of the ruled surface.

 \para

\begin{theorem}\label{T-implicharac1} A surface $\cal V$ defined implicitly by a polynomial  $f(\ox)\in {\Bbb K}[\ox]$  is a  rational  ruled surface if and only if the following statements holds:
\begin{itemize}
\item[1.]  There exist two rational  plane  curves ${\cal C}^{ij}$ and ${\cal C}^{k\ell}$ defined by a factor of $f_0^{ij}$ and $f_0^{k\ell}$, respectively,    $ij\not=k\ell$,  $ij, k\ell\in \{12,23,13\}$ (see Section 2). Let us assume that $ij=12$, and $k\ell=23$, and let $\cP^{12}=(p_1,p_2)\in {\Bbb K}(t_1)^2$, $\cP^{23}=(\tilde{p}_1,\tilde{p}_2)\in {\Bbb K}(t_1)^2$ be rational  proper parametrizations of ${\cal C}^{12}$ and ${\cal C}^{23}$, respectively.

\item[2.] If $p_1\not=0$, there exists $R(t_1), S(t_1)\in {\Bbb K}(t_1)\setminus {\Bbb K}$ such that one of the following statements holds:
\begin{itemize}
\item[2.1.]   $f({\cal P})=0,$ where
\[{\cal P}(\ot)=\left(p_1(S(t_1))-t_2\frac{p_1(S(t_1))}{\tilde{p}_2(R(t_1))}, p_2(S(t_1))+t_2\frac{\tilde{p}_1(R(t_1))-p_2(S(t_1))}{\tilde{p}_2(R(t_1))},t_2\right).\]
  In this case, ${\cal P}$ is a  rational proper parametrization of $\cal V$ in standard reduced form.
 \item[2.2.]  $f\left({\cal P}\right)=0,$ where  \[{\cal P}(\ot)=\left(p_1(S(t_1))-t_2\frac{p_1(S(t_1))}{\tilde{p}_2(R(t_1))}, p_2(S(t_1)),t_2\right).\]
 In this case, ${\cal P}$ is a rational proper parametrization of $\cal V$ in standard reduced form.
 \end{itemize}

\item[3.] If $p_1=0$, there exist  $R(t_1)\in {\Bbb K}(t_1)\setminus\{0\}$, and $S(t_1)\in {\Bbb K}(t_1)\setminus {\Bbb K}$ such that one of the following statements holds:
\begin{itemize}
\item[3.1.]   $f({\cal P})=0,$ where
$$\left\{\begin{array}{ll}
{\cal P}(\ot)=\left(t_2\frac{q_1(S(t_1))}{q_2(S(t_1))}, R(t_1)-t_2\frac{R(t_1)}{{q}_2(S(t_1))},t_2\right),&\begin{array}{l}  \mbox{if $\cP^{13}=(q_1,q_2),\,q_2\not=0$ is a}\\ \mbox{proper parametrization of  ${\cal C}^{13}$} \end{array}\\
\\
{\cal P}(\ot)=(t_2S(t_1), t_2R(t_1), t_2),\,\,R\not\in{\Bbb K}  &\begin{array}{l}  \mbox{if $\cP^{13}=(t_1,0)$ is a}\\ \mbox{ parametrization of   ${\cal C}^{13}$} \end{array}\\
\end{array}
\right.$$
  In this case, ${\cal P}$ is a  rational proper parametrization of $\cal V$ in standard reduced form.
\item[3.2.] $f({\cal P})=0,$ where
\[{\cal P}(\ot)=(t_2S(t_1), R(t_1), t_2)\in {\Bbb K}(\ot)^3,\,\,\,\,R\not\in{\Bbb K}.\]
  In this case, ${\cal P}$ is a  rational proper parametrization of $\cal V$ in standard reduced form.
   \end{itemize}
 \end{itemize}
\end{theorem}

\vspace*{2mm}

\noindent {\bf Proof.} It is clear that if statements $1$ and $2$ (or $3$) hold, then  $\cal V$ is a  rational  ruled surface. Reciprocally, let $\cal V$ be  a  rational  ruled surface. Then, a proper parametrization of $\cal V$ is given by the standard form parametrization
\[{\cal Q}(\ot)=(m_1(t_1)+t_2n_1(t_1), m_2(t_1)+t_2n_2(t_1), m_3(t_1)+t_2n_3(t_1))\in {\Bbb K}(\ot)^3.\] We  assume   that $n_1n_3\not=0$ (see Section 2, and Remark~\ref{R-ij}). Thus,
${\cal Q}(t_1, -m_1/n_1)$ parametrizes ${\cal C}^{23}$, and ${\cal Q}(t_1, -m_3/n_3)$ parametrizes ${\cal C}^{12}$. Hence, statement 1 holds.  Let us prove that statement 2  holds. For this purpose, we consider $\cP^{12}=(p_1,p_2)\in {\Bbb K}(t_1)^2$, $\cP^{23}=(\tilde{p}_1,\tilde{p}_2)\in {\Bbb K}(t_1)^2$ two rational proper parametrizations of ${\cal C}^{12}$ and ${\cal C}^{23}$, respectively. We   distinguish two different cases depending on whether  $p_1\not=0$ or $p_1=0$.
 \begin{enumerate}
\item Let $p_1\not=0$. Thus, again we distinguish two cases:
 \begin{itemize}
 \item[a.] Let $n_2\not=0$.  From the results in Section 2, we have that the surface  $\cal V$ admits a proper  parametrization in standard reduced form
  \[{\cal P}^3(\ot)=(p_{13}(t_1)+t_2q_{13}(t_1), p_{23}(t_1)+t_2q_{23}(t_1), t_2)\in {\Bbb K}(\ot)^3,\]
  such that $q_{j3}=n_j/n_3\not=0$,\,$j=1,2$ (note that $n_1n_2n_3\not=0$).   Observe that ${\cal P}^3(t_1, 0)=(p_{13},p_{23})\in {\Bbb K}(t_1)^2$ is a rational parametrization of ${\cal C}^{12}$. Then, since $\cP^{12}$ is a rational proper parametrization of  ${\cal C}^{12}$, we deduce that there exits $S\in {\Bbb K}(t_1)\setminus{\Bbb K}$ such that  $$\cP^{12}(S)=(p_1(S),p_2(S))=(p_{13},p_{23}).$$  In addition, since $\cP^{23}=(\tilde{p}_1,\tilde{p}_2)\in {\Bbb K}(t_1)^2$ is a rational proper parametrization of ${\cal C}^{23}$ and $$\cP^3(t_1, -p_{13}/q_{13})=\cP^3(t_1, -p_1(S)/q_{13})=(p_2(S)-q_{23}p_1(S)/q_{13}, -p_1(S)/q_{13})$$   is a rational  parametrization of ${\cal C}^{23}$, we deduce that there exists  $R\in {\Bbb K}(t_1)\setminus {\Bbb K}$ such that
  $$\tilde{p}_1(R)=p_2(S)-q_{23}p_1(S)/q_{13},\quad \mbox{and}\quad  \tilde{p}_2(R)=-p_1(S)/q_{13}.$$
 Note that since $p_1\not=0$ and $S, R\in   {\Bbb K}(t_1)\setminus{\Bbb K}$, then  $ \tilde{p}_2(R)\not=0$.  Then
  \[ q_{13}=\frac{-p_1(S)}{\tilde{p}_2(R)},\quad \mbox{and}\quad  q_{23}=\frac{\tilde{p}_1(R)-p_2(S)}{\tilde{p}_2(R)}.\]
Since ${\cal P}^3$ parametrizes properly $\cal V$, we   have that $f({\cal P})=0,$ where ${\cal P}$ is the proper parametrization in standard reduced form
\[{\cal P}(\ot)=\left(p_1(S(t_1))-t_2\frac{p_1(S(t_1))}{\tilde{p}_2(R(t_1))}, p_2(S(t_1))+t_2\frac{\tilde{p}_1(R(t_1))-p_2(S(t_1))}{\tilde{p}_2(R(t_1))},t_2\right).\]
 \item[b.]  Let   $n_2=0$.  From the results in Section 2, we have that the surface  $\cal V$ admits a proper parametrization in standard reduced form
\[{\cal P}^3(\ot)=(p_{13}(t_1)+t_2q_{13}(t_1),m_2(t_1), t_2),\]
such that   $q_{13}=n_1/n_3\not=0$ (note that $n_1n_3\not=0$).  Observe that ${\cal P}^3(t_1, 0)=(p_{13},m_2)\in {\Bbb K}(t_1)^2$ is a rational parametrization of ${\cal C}^{12}$. Then, since $\cP^{12}$ is a rational proper parametrization of  ${\cal C}^{12}$, there exist $S\in {\Bbb K}(t_1)\setminus{\Bbb K}$ such that   $$\cP^{12}(S)=(p_1(S),p_2(S))=(p_{13},m_2).$$  In addition, since $\cP^{23}=(\tilde{p}_1,\tilde{p}_2)\in {\Bbb K}(t_1)^2$ is a rational parametrization of ${\cal C}^{23}$ and $$\cP^3(t_1, -p_{13}/q_{13})=\cP^3(t_1, -p_1(S)/q_{13})=(p_2(S), -p_1(S)/q_{13})$$   is a rational parametrization of ${\cal C}^{23}$, we deduce that there exists  $R\in {\Bbb K}(t_1)\setminus {\Bbb K}$ such that
 $$\tilde{p}_1(R)=p_2(S),\quad \mbox{and}\quad \tilde{p}_2(R)=-p_1(S)/q_{13}.$$
  Note that since $p_1\not=0$ and $S, R\in   {\Bbb K}(t_1)\setminus{\Bbb K}$, then  $\tilde{p}_2(R)\not=0$.   Hence,
 $$p_2(S)=\tilde{p}_1(R),\quad \mbox{and}\quad q_{13}=\frac{-p_1(S)}{\tilde{p}_2(R)}.$$
 Since ${\cal P}^3$ parametrizes properly $\cal V$, we  have that  $f\left({\cal P}\right)=0,$ where ${\cal P}$ is the proper parametrization in standard reduced form  \[{\cal P}(\ot)=\left(p_1(S(t_1))-t_2\frac{p_1(S(t_1))}{\tilde{p}_2(R(t_1))}, p_2(S(t_1)),t_2\right).\]
\end{itemize}
\item Let $p_1=0$.   Observe that from the above proof, we deduce that $\tilde{p}_2=p_{13}=0$, and  the surface  $\cal V$ admits a proper parametrization of the form
  \[{\cal P}^3(\ot)=(t_2q_{13}(t_1), p_{23}(t_1)+t_2q_{23}(t_1), t_2)\in {\Bbb K}(\ot)^3,\quad q_{i3}=\frac{n_i}{n_3},\,i=1,2,\,\,q_{13}\not=0.\]
  Under these conditions, we distinguish two different cases.
  \begin{itemize}
 \item[a.] Let $n_2\not=0$.  Then $q_{23}=n_2/n_3\not=0$, and ${\cal P}^3(t_1, -p_{23}/q_{23})=(-q_{13}p_{23}/q_{23}, -p_{23}/q_{23})\in {\Bbb K}(t_1)^2$ is a rational parametrization of ${\cal C}^{13}$. Let  $\cP^{13}=(q_1, q_2)$ be  a rational proper parametrization of  ${\cal C}^{13}$.  Thus, there exits $S\in {\Bbb K}(t_1)\setminus{\Bbb K}$ such that  $$\cP^{13}(S)=(q_1(S),q_2(S))=(-q_{13}p_{23}/q_{23}, -p_{23}/q_{23}).$$
     \begin{itemize}
 \item[a.1.]   If $q_2\not=0$, since $S\not\in {\Bbb K}$ we get that $q_2(S)\not=0$, and then  \[ q_{13}=\frac{q_1(S)}{q_2(S)},\quad \mbox{and}\quad q_{23}=-\frac{p_{23}}{{q}_2(S)}.\]
Since ${\cal P}^3$ parametrizes properly $\cal V$, we   have that $f({\cal P})=0,$ where ${\cal P}$ is the proper parametrization  in standard reduced form
\[{\cal P}(\ot)=\left(t_2\frac{q_1(S(t_1))}{q_2(S(t_1))},  R(t_1)-t_2\frac{R(t_1)}{{q}_2(S(t_1))},t_2\right),\quad R(t_1):=p_{23}(t_1).\]
Note that $R\not=0$ because $\cal V$ is not   the plane $x_2=0$ (see Section 2).
  \item[a.2.] If $q_2=0$, then  $p_{23}=0$  and
  since ${\cal P}^3$ parametrizes properly $\cal V$, we   have that $f({\cal P})=0,$ where ${\cal P}$ is the proper parametrization  in standard reduced form
\[{\cal P}(\ot)=(t_2S(t_1), t_2R(t_1), t_2)\in {\Bbb K}(\ot)^3,\quad S:=q_{13},\,R:=q_{23}.\]
Note that $S, R\not\in {\Bbb K}$ because $\cal V$ is not a cylinder over the coordinate planes (see Section 2).
   \end{itemize}
 \item[b.] If $n_2=0$.  Then $q_{23}=n_2/n_3=0$, and  since ${\cal P}^3$ parametrizes properly $\cal V$, we   have that $f({\cal P})=0,$ where ${\cal P}$ is the proper parametrization  in standard reduced form
\[{\cal P}(\ot)=(t_2S(t_1), R(t_1), t_2)\in {\Bbb K}(\ot)^3,\quad S:=q_{13},\,R:=p_{23}.\]
Note that $S, R\not\in {\Bbb K}$ because $\cal V$ is not a cylinder over the coordinate planes, and $\cal V$ is not the plane $x_2-c=0,\,c\in {\Bbb K}$ (see Section 2).
 \end{itemize}
 \end{enumerate}
 \qed

\begin{remark}\label{R-ij}
%Note that in Theorem~\ref{T-implicharac1}, we assume that ${\cal C}^{12}$ and ${\cal C}^{23}$ are the two rational curves satisfying statement~1. In addition, in the proof of theorem, when a ruled surface is given, we consider a parametrization $\cal Q$ in standard form (see~\ref{eq-standardform})
%%\[{\cal Q}(\ot)=(m_1(t_1)+t_2n_1(t_1), m_2(t_1)+t_2n_2(t_1), m_3(t_1)+t_2n_3(t_1))\in {\Bbb K}(\ot)^3\] a rational proper parametrization
%such that  $n_1n_3\not=0$ (from this assumption, we have that
%${\cal Q}(t_1, -m_1/n_1)$ parametrizes ${\cal C}^{23}$, and ${\cal Q}(t_1, -m_3/n_3)$ parametrizes ${\cal C}^{12}$).\\
Theorem~\ref{T-implicharac1} can be proved similarly if a different pair of rational  plane  curves ${\cal C}^{ij}$ are considered in statement~1, and if a different pair of polynomials $n_i, n_j$ satisfies that  $n_in_j\not=0$ (see Section 2).\\

\noindent
In the following, we assume that $n_1n_3\not=0$, and that ${\cal C}^{12}$ and ${\cal C}^{23}$ are the two rational  plane  curves satisfying statement~1 in Theorem~\ref{T-implicharac1}.  This requirement can always be achieved by applying a linear
transformation to $\cal V$,  and therefore it is not a loss of
generality for our purposes since one can always undo the linear
transformation once    $\cal P$ has been computed.
\end{remark}

\para

In Corollary~\ref{C-implicharac1}, we prove that statements 2 and 3 in Theorem~\ref{T-implicharac1} are equivalent to check the rationality of a plane curve, and to compute, in the affirmative case, a rational parametrization. For this purpose,  we use the notion  of content and primitive part of a polynomial.  More precisely, given
a non-zero polynomial $a(x_1,\ldots,x_n)\in I[x_1,\ldots,x_n]$, where   $I$ is a unique
factorization domain, the content of  $a$ w.r.t $\ox:=(x_1,\ldots, x_j),\,j\leq n$  is the gcd
of all the coefficients of $a(\ox)$ with respect to $\ox$.  We denote it by $\Content_{\ox}(a)$. Observe that  $\Content_{\ox}(a)$ divides the polynomial $a$. In addition, we denote by $\pp_{\ox}(a)$ the primitive
part of $a$ w.r.t. $\ox$. We have that
$a(\ox)=\Content_{\ox}(a)\,\pp_{\ox}(a)$, and it holds that   the gcd of all coefficients of
$\pp_{\ox}(a)$ is 1  (see \cite{win}).

\para

  Finally, using the notation introduced in Theorem~\ref{T-implicharac1}, we consider the polynomials    $N_i(x_1,x_2)=\content_{t_2}(g_i),\,\,i\in\{1,\ldots, 5\}$, where
  %$N_i(x,y)=\content_{t_2}(g_i)/(\content_{x}(g_i)\content_{y}(g_i)),\,\,i\in\{1,\ldots, 5\}$, where
\[g_1(x_1,x_2,t_2)=\numer\left(f\left(p_1(x_2)-t_2\frac{p_1(x_2)}{\tilde{p}_2(x_1)}, p_2(x_2)+t_2\frac{\tilde{p}_1(x_1)-p_2(x_2)}{\tilde{p}_2(x_1)},t_2\right)\right),\]
\[g_2(x_1,x_2,t_2)=\numer\left(f\left(p_1(x_2)-t_2\frac{p_1(x_2)}{\tilde{p}_2(x_1)}, p_2(x_2),t_2\right)\right),\]
\[g_3(x_1,x_2,t_2)=\numer\left(f\left(t_2\frac{q_1(x_2)}{q_2(x_2)},  x_1-t_2\frac{x_1}{{q}_2(x_2)},t_2\right)\right),\]\[g_4(x_1,x_2,t_2)=f\left(t_2x_2, t_2x_1, t_2\right),\quad \mbox{and}\quad g_5(x_1,x_2,t_2)=f\left(t_2x_2, x_1, t_2\right).\]

\para

Observe that $g_i(x_1,x_2,t_2)=N_i(x_1,x_2)M_i(x_1,x_2,t_2)$, where $M_i(x_1,x_2,t_2)=\pp_{t_2}(g_i)$, for $i\in\{1,\ldots, 5\}$.

\para

\noindent
Under these conditions, we prove the following corollary.

\para

\begin{corollary}\label{C-implicharac1}  Let $\cal V$ be a surface defined implicitly by a polynomial  $f(\ox)\in {\Bbb K}[\ox]$ and such that statement~1 in Theorem~\ref{T-implicharac1} holds. $\cal V$ is a  rational  ruled surface if and only if for some $i\in\{1,\ldots, 5\}$, there exists a factor of $N_i$ defining a rational  plane  curve ${\cal C}_{N_i}$. In this case, $(R(t_1), S(t_1))\in {\Bbb K}(t_1)^2$, where $S\not\in {\Bbb K}$, is a rational parametrization of ${\cal C}_{N_i}$.
\end{corollary}

\vspace*{2mm}

\noindent {\bf Proof.} Let us prove that statement $2.1$ in Theorem~\ref{T-implicharac1} is equivalent to the existence of a factor of $N_1$ defining a rational  plane  curve ${\cal C}_{N_1}$. For this purpose, we write
\[g_1(x_1,x_2, t_2)=h_0(x_1,x_2)+h_1(x_1,x_2)t_2+\cdots+h_n(x_1,x_2)t_2^n.\]
Observe that since \[g_1(x_1,x_2,t_2)=\numer\left(f\left(p_1(x_2)-t_2\frac{p_1(x_2)}{\tilde{p}_2(x_1)}, p_2(x_2)+t_2\frac{\tilde{p}_1(x_1)-p_2(x_2)}{\tilde{p}_2(x_1)},t_2\right)\right),\] and $f(p_1(x_2),p_2(x_2),0)=0$, then $t_2$ divides $g_1$. Thus, $h_0=0$.\\
First, we note that if $f({\cal P})=0,$ where
\[{\cal P}(\ot)=\left(p_1(S(t_1))-t_2\frac{p_1(S(t_1))}{\tilde{p}_2(R(t_1))}, p_2(S(t_1))+t_2\frac{\tilde{p}_1(R(t_1))-p_2(S(t_1))}{\tilde{p}_2(R(t_1))},t_2\right),\]
then $(R(t_1), S(t_1))\in ({\Bbb K}(t_1)\setminus{\Bbb K})^2$ is such that $g_1(R(t_1), S(t_1),t_2)=0$. Since  $(R(t_1), S(t_1))$ does not depend on $t_2$, we get that $h_j(R, S)=0,\,j=1,\ldots,n$. Then, $h(x_1,x_2)$ divides $N_1(x_1,x_2)=\gcd(h_1,\ldots, h_n)$, where $h$ is the implicit polynomial defining the curve parametrized by $(R,S)$. Therefore, $N_1(R, S)=0$ and      $(R, S)$ is a rational parametrization of the plane curve  ${\cal C}_{N_1}$  defined by a factor of the polynomial $N_1.$\\
Reciprocally, let ${\cal Q}(t_1)=(U(t_1),V(t_1))\in {\Bbb K}(t_1)^2$ be such that $N_1({\cal Q})=0$. Since $N_1$ divides $g_1$, we deduce that $g_1(U(t_1),V(t_1), t_2)=0$. Hence, $f({\cal P})=0,$ where
\[{\cal P}(\ot)=\left(p_1(V(t_1))-t_2\frac{p_1(V(t_1))}{\tilde{p}_2(U(t_1))}, p_2(V(t_1))+t_2\frac{\tilde{p}_1(U(t_1))-p_2(V(t_1))}{\tilde{p}_2(U(t_1))},t_2\right).\]
One reasons similarly to prove that statement $2.2$ in Theorem~\ref{T-implicharac1} is equivalent to the existence of a factor of $N_2$ defining a rational  plane  curve ${\cal C}_{N_2}$, that statement $3.1$ in Theorem~\ref{T-implicharac1} is equivalent to the existence of a factor of $N_3$ or $N_4$ defining a rational  plane  curve ${\cal C}_{N_3}$ or ${\cal C}_{N_4}$, and that    statement $3.2$ in Theorem~\ref{T-implicharac1} is equivalent to the existence of a factor of $N_5$ defining a rational  plane  curve ${\cal C}_{N_5}$.
 \qed

\para

From  Theorem~\ref{T-implicharac1} and Corollary~\ref{C-implicharac1}, we derive the following algorithm that decides whether $\cal V$ is a  rational  ruled surface and in the affirmative case, it computes  a rational proper  parametrization of $\cal V$.

\para

\noindent
\fbox{{\sf Algorithm Parametrization of a Rational Ruled Surface 2.}}
\begin{itemize}
\item {\bf Input:} A surface $\cal V$ defined by an irreducible polynomial $f(\ox)\in {\Bbb K}[\ox]$.
\item {\bf Output:} the message ``{\tt $\cal V$ is not a  rational  ruled surface}" or
a proper parametrization ${\cal P}$ of ``{\tt the  rational  ruled surface $\cal V$}".
\end{itemize}
 \begin{itemize}
 \item[1.] If $\deg_{x_3}(f)=0$ (similarly if  $\deg_{x_1}(f)=0$  or $\deg_{x_2}(f)=0$), compute  $(p(t_1), q(t_1))$ a parametrization of the curve defined by the polynomial $f(x_1, x_2)=0$, and  {\sc Return} $\cP(\ot)=(p(t_1), q(t_1), t_2)\in {\Bbb K}(\ot)^3$
 ``{\tt  is a proper parametrization of the  rational  ruled surface  ${\cal V}$  in stan\-dard re\-du\-ced form}".
  \item[2.] Compute the polynomials $f_{0}^{ij}(x_i, x_j)$, and check whether there exist two rational  plane  curves ${\cal C}^{ij}$ and ${\cal C}^{k\ell}$ defined by a factor of $f_0^{ij}$ and $f_0^{k\ell}$, respectively,  for $ij\not=k\ell$, and $ij, k\ell\in \{12,23,13\}$ . In the affirmative case,  we assume that $ij=12$, and $k\ell=23$ (see Remark~\ref{R-ij}). Otherwise, {\sc Return} ``{\tt $\cal V$ is not a  rational  ruled surface}".
 \item[3.] Compute $\cP^{12}=(p_1,p_2)\in {\Bbb K}(t_1)^2$, and $\cP^{23}=(\tilde{p}_1,\tilde{p}_2)\in {\Bbb K}(t_1)^2$ rational proper parametrizations of ${\cal C}^{12}$ and ${\cal C}^{23}$, respectively. For this purpose, apply for instance the results in Sections 4.7 and 4.8 in \cite{libro}.   If $p_1\not=0$ go to Step 4. Otherwise, go to Step 6.
     \item[4.] Check whether there exists a rational  plane  curve ${\cal C}_{N_1}$ defined by a factor of the polynomial $N_1(x_1,x_2)= \content_{t_2}(g_1)$, where
\[g_1(x_1,x_2,t_2)=\numer\left(f\left(p_1(x_2)-t_2\frac{p_1(x_2)}{\tilde{p}_2(x_1)}, p_2(x_2)+t_2\frac{\tilde{p}_1(x_1)-p_2(x_2)}{\tilde{p}_2(x_1)},t_2\right)\right).\]
In the affirmative  case, compute $(R(t_1), S(t_1))\in ({\Bbb K}(t_1)\setminus{\Bbb K})^2$   a rational proper parametrization of
${\cal C}_{N_1}$, and  {\sc Return} \[{\cal P}(\ot)=\left(p_1(S(t_1))-t_2\frac{p_1(S(t_1))}{\tilde{p}_2(R(t_1))}, p_2(S(t_1))+t_2\frac{\tilde{p}_1(R(t_1))-p_2(S(t_1))}{\tilde{p}_2(R(t_1))},t_2\right),\]
    ``{\tt  is a proper parametrization of the  rational  ruled surface  ${\cal V}$  in stan\-dard re\-du\-ced form}". Otherwise, go to Step 5.
\item[5.] Check whether there exists a rational  plane  curve ${\cal C}_{N_2}$ defined by a factor of the polynomial
 $N_2(x_1,x_2)=\content_{t_2}(g_2)$, where
\[g_2(x_1,x_2,t_2)=\numer\left(f\left(p_1(x_2)-t_2\frac{p_1(x_2)}{\tilde{p}_2(x_1)}, p_2(x_2),t_2\right)\right).\]
 In the affirmative  case, compute $(R(t_1), S(t_1))\in ({\Bbb K}(t_1)\setminus{\Bbb K})^2$   a rational proper parametrization of
${\cal C}_{N_2}$, and  {\sc Return}
\[{\cal P}(\ot)=\left(p_1(S(t_1))-t_2\frac{p_1(S(t_1))}{\tilde{p}_2(R(t_1))}, p_2(S(t_1)),t_2\right)\]
   ``{\tt  is a proper parametrization of the  rational  ruled surface  ${\cal V}$  in stan\-dard re\-du\-ced form}". Otherwise, go to Step 2, and consider different rational components and apply again the algorithm. If there have no more rational components, {\sc Return} ``{\tt $\cal V$ is not a  rational  ruled surface}".
   \item[6.]  Check whether the plane curve ${\cal C}^{13}$ is rational. In the affirmative case, compute   $\cP^{13}=(q_1,q_2)\in {\Bbb K}(t_1)^2$   a rational  proper parametrization  of ${\cal C}^{13}$ (apply for instance the results in Sections 4.7 and 4.8 in \cite{libro}).  Otherwise, go to Step 8.
            \item[7.]
            \begin{itemize}
                \item[7.1.]         If $q_2\not=0$, check whether there exists a rational  plane  curve ${\cal C}_{N_3}$ defined by a factor of the polynomial
  $N_3(x_1,x_2)=\content_{t_2}(g_3)$, where
\[g_3(x_1,x_2,t_2)=\numer\left(f\left(t_2\frac{q_1(x_2)}{q_2(x_2)},  x_1-t_2\frac{x_1}{{q}_2(x_2)},t_2\right)\right). \]
In the affirmative  case, compute $(R(t_1), S(t_1))\in {\Bbb K}(t_1)^2$,\,$R\not=0,\,S\not\in {\Bbb K}$,  a rational proper parametrization of
${\cal C}_{N_3}$   and  {\sc Return} \[{\cal P}(\ot)=\left(t_2\frac{q_1(S(t_1))}{q_2(S(t_1))},  R(t_1)-t_2\frac{R(t_1)}{{q}_2(S(t_1))},t_2\right),\]
    ``{\tt  is a proper parametrization of the  rational  ruled surface  ${\cal V}$ in stan\-dard reduced form}". Otherwise, go to Step 2, and consider different rational components and apply again the algorithm. If there have no more rational components, {\sc Return} ``{\tt $\cal V$ is not a  rational  ruled surface}".
                    \item[7.2.]      If $q_2=0$, check whether there exists a rational  plane  curve ${\cal C}_{N_4}$ defined by a factor of the polynomial
  $N_4(x_1,x_2)=\content_{t_2}(g_4)$, where
\[g_4(x_1,x_2,t_2)=f\left(t_2x_2, t_2x_1, t_2\right). \]
  In the affirmative  case, compute $(R(t_1), S(t_1))\in ({\Bbb K}(t_1)\setminus{\Bbb K})^2$   a rational proper parametrization of
${\cal C}_{N_4}$   and  {\sc Return} \[{\cal P}(\ot)=\left(t_2S(t_1),  t_2R(t_1),t_2\right),\]
    ``{\tt  is a proper parametrization of the  rational  ruled surface  ${\cal V}$  in stan\-dard re\-du\-ced form}".  Otherwise, go to Step 2, and consider different rational components and apply again the algorithm. If there have no more rational components, {\sc Return} ``{\tt $\cal V$ is not a  rational  ruled surface}".
      \end{itemize}
                   \item[8.] Check  whether there exists a rational  plane  curve ${\cal C}_{N_5}$ defined by a factor of the polynomial
  $N_5(x_1,x_2)=\content_{t_2}(g_5)$, where
\[g_5(x_1,x_2,t_2)=f\left(t_2x_2, x_1, t_2\right). \]
  In the affirmative  case, compute $(R(t_1), S(t_1))\in ({\Bbb K}(t_1)\setminus{\Bbb K})^2$  a rational proper parametrization of
${\cal C}_{N_5}$   and  {\sc Return} \[{\cal P}(\ot)=\left(t_2S(t_1),  R(t_1),t_2\right),\]
    ``{\tt  is a proper parametrization of the  rational  ruled surface  ${\cal V}$  in stan\-dard re\-du\-ced form}". Otherwise, go to Step 2, and consider different rational components and apply again the algorithm. If there have no more rational components, {\sc Return} ``{\tt $\cal V$ is not a  rational  ruled surface}".
  \end{itemize}

 \begin{remark}\label{R-real1}  Remark~\ref{R-Alg2} can be stated similarly in this new situation to   the rational parametrizations $\cP^{12}, \cP^{23}, \cP^{13}$, $(R,S)$, and the output parametrization $\cal P$.
\end{remark}

\para

In the following example, we illustrate the performance of  {\sf Algorithm Parametrization of a Rational Ruled Surface 2}.

\para

\begin{example} Consider the surface $\cal V$ over $\Bbb C$ introduced in Example 2, and defined implicitly by the polynomial \\

\noindent
$f(\ox)=-49x_2x_1^3-799x_3x_2x_1^2+20x_2x_1^2+2x_2^2x_1^2+980x_3x_1^2-2205x_3^2x_1^2+x_2^3x_1-33750x_3^3x_1-400x_3x_1+606x_3x_2x_1-5x_2^2x_1-68x_3x_2^2x_1-1747x_3^2x_2x_1-25x_3^2x_1+x_2^3x_3-25x_2^2x_3^2+1396x_2x_3^2-1120x_3^2-48915x_3^4-5190x_3^3-4237x_2x_3^3-14x_2^2x_3\in {\Bbb C}[\ox].$\\

\noindent
Let us apply {\sf Algorithm Parametrization of a Rational Ruled Surface 2}. For this purpose, we first observe $\deg_{x_i}(f)>0$, for $i=1,2,3$. Then, in Step 2 of the algorithm, we compute  $$f_0^{12}(x_1,x_2)=-x_1x_2(49x_1^2-20x_1-2x_1x_2-x_2^2+5x_2),\quad $$$$ f_0^{23}(x_2,x_3)=(14x_2^2+5190x_3^2+1120x_3+48915x_3^3-1396x_2x_3+25x_2^2x_3+4237x_2x_3^2-x_2^3).$$ The plane curves defined by the equations $49x_1^2-20x_1-2x_1x_2-x_2^2+5x_2=0$, and $14x_2^2+5190x_3^2+1120x_3+48915x_3^3-1396x_2x_3+25x_2^2x_3+4237x_2x_3^2-x_2^3=0$ are rational. These curves are denoted as ${\cal C}^{12}$ and ${\cal C}^{23}$, respectively.\\

\noindent
In Step 3 of the algorithm, we  compute a rational proper parametrization of ${\cal C}^{12}$,
\[\cP^{12}(t_1)=(p_1,p_2)=\left(\frac{-\sqrt{2}t_1(-5+t_1)}{5(4t_1-10+5\sqrt{2})}, \frac{(50+5\sqrt{2})t_1(-20+100\sqrt{2}+49t_1)}{1225(4t_1-10+5\sqrt{2})}\right)\in {\Bbb R}(t_1)^2.\]
and  ${\cal C}^{23}$,
$$\cP^{23}(t_1)=(\tilde{p}_1(t_1),\tilde{p}_2(t_1))=\left(\frac{2 (-378367 t_1^2+10410900 t_1-142098075+4102 t_1^3)}{1241(t_1^3-25 t_1^2-4237 t_1-48915)}, \right.$$$$ \left.\frac{2(20322550-513355 t_1-28 t_1^2+49 t_1^3)}{1241(t_1^3-25 t_1^2-4237 t_1-48915)}\right)\in {\Bbb R}(t_1)^2.$$
Since   $p_1\not=0$ we go to Step 4 of the algorithm, and we check whether there exists a rational  plane  curve ${\cal C}_{N_1}$ defined by a factor of the polynomial $N_1(x_1,x_2)= \content_{t_2}(g_1)$, where
\[g_1(x_1,x_2,t_2)=\numer\left(f\left(p_1(x_2)-t_2\frac{p_1(x_2)}{\tilde{p}_2(x_1)}, p_2(x_2)+t_2\frac{\tilde{p}_1(x_1)-p_2(x_2)}{\tilde{p}_2(x_1)},t_2\right)\right).\]
We have that\\

\noindent
$N_1(x_1, x_2)=(73 x_2 x_1-2555 x_2+2482 x_2\sqrt{2}+18675-4150\sqrt{2}-315 x_1+70\sqrt{2} x_1)(4x_2-10+5\sqrt{2})(x_1-35+34\sqrt{2})^3(-x_1+35+34\sqrt{2})^3(x_1+45)^3.$\\

\noindent
Since we look for $(R,S)\in ({\Bbb C}(t_1)\setminus{\Bbb C})^2$, we consider the plane curve defined by the irreducible polynomial $73 x_2 x_1-2555 x_2+2482 x_2\sqrt{2}+18675-4150\sqrt{2}-315 x_1+70\sqrt{2} x_1$, and we observe that it  defines a rational  plane  curve ${\cal C}_{N_1}$. Then, we compute a rational proper parametrization of ${\cal C}_{N_1}$. We obtain
 \[(R(t_1), S(t_1))=\left(t_1, -\frac{5(-9+2\sqrt{2})(-415+7t_1)}{73(t_1-35+34\sqrt{2})}\right)\in ({\Bbb R}(t_1)\setminus{\Bbb R})^2.\]
Then,
 \[{\cal P}=\left(p_1(S(t_1))-t_2\frac{p_1(S(t_1))}{\tilde{p}_2(R(t_1))}, p_2(S(t_1))+t_2\frac{\tilde{p}_1(R(t_1))-p_2(S(t_1))}{\tilde{p}_2(R(t_1))},t_2\right)=\left(\frac{q_{11}}{q_{12}},\frac{q_{21}}{q_{22}}, t_2\right),\]
where\\

\noindent
$q_{11}=(-9+2 \sqrt{2}) \sqrt{2} (-490 t_1^3+280 t_1^2+5133550 t_1-203225500+392 t_1^2 \sqrt{2}+7186970 \sqrt{2} t_1-284515700 \sqrt{2}-686 t_1^3 \sqrt{2}+6205 t_2 t_1^3+8687 t_2 \sqrt{2} t_1^3-155125 t_2 t_1^2-217175 t_2 t_1^2 \sqrt{2}-26290585 t_2 t_1-36806819 t_2 \sqrt{2} t_1-303517575 t_2-424924605 t_2 \sqrt{2}),$\\

\noindent
$q_{12}= 73(-415+7 t_1) (-4866+106 t_1-4199 \sqrt{2}+17 \sqrt{2} t_1) (t_1-35+34 \sqrt{2}),$\\

\noindent
$q_{21}=2068272 t_1^2-544956812 t_1+331704 t_1^2 \sqrt{2}-87398734 \sqrt{2} t_1+63812 t_1^3+10234 t_1^3 \sqrt{2}+11741634840+1883092380 \sqrt{2}+78183 t_2 \sqrt{2} t_1^3-10107945 t_2 t_1^2 \sqrt{2}+239474529 t_2 \sqrt{2} t_1+487494 t_2 t_1^3-63026010 t_2 t_1^2+1493194122 t_2 t_1+5038391745 t_2 \sqrt{2}+31415854410 t_2$,\\

\noindent
$q_{22}=146(t_1+118) (-4866+106 t_1-4199 \sqrt{2}+17 \sqrt{2} t_1) (t_1-35+34 \sqrt{2})$\\

\noindent
    ``{\tt  is a proper parametrization of the  rational  ruled surface  ${\cal V}$  in standard reduced form}".\\

 \noindent
    Observe that since $\cP^{12},\,\cP^{23}$ and $R, S$ have coefficients in $\Bbb R$, then ${\cal P}$ also has coefficients in  $\Bbb R$
 (see Remark~\ref{R-real1}).
      \end{example}

%----------------------------------------------------------------

\section{Parametrically Ruled Surfaces}

In this section, we consider a surface $\cal V$ defined over an algebraically closed field of characteristic zero $\Bbb K$, and a parametrization (not necessarily proper) that defines   $ \cal V$,
 \[{\cal M}(\ot)=(m_1(\ot), m_2(\ot), m_3(\ot))\in {\Bbb K}(\ot)^3.\]

In this section, we analyze whether $\cal V$ is a ruled surface, and in the affirmative case we compute a proper reparametrization in standard reduced form (see equation given in~\ref{eq-standardreducedform}). More precisely, the idea is to check whether there exists a parametrization of the form
\[{\cal P}(\ot)=(p_1(t_1)+t_2q_1(t_1), p_2(t_1)+t_2q_2(t_1), t_2)\in {\Bbb K}(\ot)^3,\]
(where $p_j, q_j$ are given in Theorem~\ref{T-implicharac1}), and $(U, V)\in ({\Bbb K}(\ot)\setminus{\Bbb K})^2$ such that ${\cal P}(U, V)={\cal M}$. Observe that from this equality, we get that $V=m_3$.\\

To start with the problem, we first need to deal with the plane case and the cylinder case. If $\cal V$ is the plane $x_1-c=0,\,c\in {\Bbb C}$ (similarly, if $\cal V$ is the plane $x_i-c=0,\,c\in {\Bbb C}$,\,$i=2,3$), then $m_1=c$, and a proper parametrization of $\cal V$ is given by $\cP(\ot)=(c,t_1, t_2)$.

\para

In order to analyze whether  $\cal V$ is a cylinder  over any of the
coordinate planes of ${\Bbb K}^3$, we apply the following result presented in \cite{sonia08} (Theorem 5).

\para

\begin{theorem} \label{T-cyl-test} {\sf (Cylinder criterion)} Let $H_i(\ot,\oh)=\numer(m_{i}(\ot)-m_{i}(\oh))$,
 where $\oh=(h_1,h_2)$, and $i\in\{1,2,3\}$. Then,
$\mathcal V$ is a cylinder over the $x_ix_j$--plane if and only if
$\gcd(H_i,H_j)\neq 1$.
\end{theorem}

\para

If $\mathcal V$ is a cylinder over the $x_1x_2$-plane and $f(x_1,x_2)\in {\Bbb K}[x_1, x_2]$ is the implicit equation defining $\cal V$,  we consider   $a\in \mathbb  K$ such that $(m_1, m_2)(a,t_2)\not\in {\Bbb K}^2$, and we get that, up to multiplication by non-zero constants,
\[f(x_1,x_2)^{r}=\Resultant_{t_2}(G_1(a,t_2,x_2), G_2(a,t_2,x_2)),\quad \mbox{where}\quad r\in {\Bbb K} \quad \mbox{and}\]
$G_i(\ot, x_i)=\numer(m_{i}(\ot)-x_{i}),\,i=1,2$ (see Theorem 8 in \cite{sonia08}). Then, one computes a proper parametrization $(p(t_1), q(t_1))\in {\Bbb K}(t_1)^2$ of the plane curve defined by the equation $f(x_1, x_2)=0$, and we get that $\cP(\ot)=(p(t_1),q(t_1), t_2)$ is a proper parametrization of $\cal V$. One reasons similarly if $\cal V$ is a cylinder over a different plane.

\para

Once the plane case and the cylinder case are analyzed, we assume that $\cal V$ is neither a cylinder nor a plane. As we stated above, we are interested in applying Theorem~\ref{T-implicharac1}. For this purpose, first we need to compute a rational proper parametrization  of ${\cal C}^{12}$ and  ${\cal C}^{23}$ (see statement 1 of Theorem~\ref{T-implicharac1}). We also need to determine a rational proper parametrization  of ${\cal C}^{13}$, if we are in statement 3 of Theorem~\ref{T-implicharac1}.

\para

Since we do not have the implicit equation defining the surface $\cal V$, we have to compute the polynomial  $f_0^{ij}(x_i,x_j)$ defining implicitly the plane curve ${\cal C}^{ij}$, $i<j$,\,$i,j\in \{1,2,3\}$, using the input parametrization $\cal M$. For this purpose, we use Theorem 10 in~\cite{sonia08}, and the fact that if $\ot_0\in {\Bbb K}^2$ is such that $m_i(\ot_0)-x_1=m_j(\ot_0)-x_2=m_k(\ot_0)=0$, for $k\in \{1,2,3\},\,k\not=i,\,k\not=j$, then $(x_1, x_2)\in {\cal C}^{ij}$. For this purpose, in order to apply Theorem 10 in~\cite{sonia08}, we need assume that none of the projective curves defined by each  numerator and denominator of
$m_i,\,i=1,2,3$   passes through the points at infinity
$(0:1:0)$ and $(1:0:0)$, where the homogeneous variables are $(t_1,t_2, w)$. Note
that this requirement can always be achieved by applying a linear change of variables
 to $\cal M$. This assumption  implies that  each numerator and denominator of
$m_i$   has positive degree w.r.t. $t_i$, and then its
leading coefficient w.r.t. $t_i$ does not depend on $t_j$,\,$i\not=j,\,i,j\in {1,2}$. Thus, for $k=1,2,3$, and $i\not=j,\,i,j\in {1,2}$,  $\deg_{t_i}(G_k(\ot,x_k))>0$, and   the leading coefficient of $G_k(\ot,x_k)$ w.r.t.
$t_i$ does not depend of $t_j$, where $G_k(\ot, x_i)=\numer(m_{k}(\ot)-x_{k}).$

\para

Finally,  for $i,j\in
\{1,2,3\}$, with $i<j$, we need to assume that the gradients $\{\nabla m_i(\ot),\,\nabla m_j(\ot) \}$ are linearly
independent as vectors in  ${\mathbb  K}(\ot)^2$. Since
$\cal M$ parametrizes a surface, the gradients of at least two of
its components  are linearly independent. In fact, since $\cal V$ is neither a cylinder nor a plane, we may assume that for every $i,j\in
\{1,2,3\}$, with $i<j$, the gradients $\{\nabla m_i(\ot),\,\nabla m_j(\ot) \}$ are linearly
independent. Note
that this requirement can always be achieved by applying a linear
transformation to $\cal M$,  and therefore it is not a loss of
generality for our purposes since one can always undo the linear
transformation once the implicit equation $f_0^{ij}$ has been computed.

  \para

 Under these conditions,  Theorem~\ref{T-impli} shows how to compute the polynomials   $f_0^{ij}(x_i,x_j)$. The theorem is obtained from Lemmas 12, 13, 14, 15, and Theorem 10 in \cite{sonia08}.

  \para

\begin{theorem}\label{T-impli} It holds that for $j<k$,\,$i\not=j,\,i\not=k$, and $i,j,k\in \{1,2,3\}$,
\[(f_0^{ij}(x_j,x_k))^{r}=\pp_{x_{k}}(\Content_{\{Z, W\}}(\Resultant_{t_2}(T(t_2, x_j),\, K(t_2, Z, W, x_j,x_k))))\in {\mathbb
K}[x_j,x_k],\,\,\]
where $ r\in {\Bbb K}$, and
\begin{enumerate}
\item $K(t_2, Z, W, x_j,x_k)=\Resultant_{t_1}(S(t_1, x_j),\, G_{Z,W}(\ot, Z, W, x_j,x_k)),$
\item $G_{Z,W}(\ot, Z, W, x_j,x_k)=G_{k}(\ot, x_{k})+ZG_i(\ot, 0)+WG_j(\ot, x_j)$,
\item $S(t_1,x_j)=\pp_{x_j}(\Resultant_{t_2}(G_i(\ot,0), G_j(\ot,x_j))),$  \item $ T(t_2, x_j
)=\pp_{x_j}(\Resultant_{t_1}(G_i(\ot,0), G_j(\ot,x_j)))$.
\end{enumerate}
\end{theorem}

 %\vspace*{2mm}

 %\noindent {\bf Proof.} We prove the theorem for  $i=1, j=2$ and $k=3$. That is,
 %\[(g_0^{ij}(x_2,x_3))^{r}=\pp_{x_{3}}(\Content_{\{Z, W\}}(\Resultant_{t_2}(T(t_2, x_2),\, K(t_2, Z, W, x_2,x_3))))\]
%where
%\begin{enumerate}
%\item $K(t_2, Z, W, x_2,x_3)=\Resultant_{t_1}(S(t_1, x_2),\, G_{Z,W}(\ot, Z, W, x_2,x_3)),$
%\item $G_{Z,W}(\ot, Z, W, x_2,x_3)=G_{3}(\ot, x_{3})+ZG_1(\ot, 0)+WG_2(\ot, x_2)$,
%\item $S(t_1,x_2)=\pp_{x_2}(\Resultant_{t_2}(G_1(\ot,0), G_2(\ot,x_2)),$  \item $ T(t_2, x_2
%)=\pp_{x_2}(\Resultant_{t_1}(G_1(\ot,0), G_2(\ot,x_2)))$.
%\end{enumerate}
% One reasons similarly for the other cases.
 %\qed

\para

Under these conditions, we apply Theorem~\ref{T-implicharac1}, and we obtain a procedure that computes a proper reparametrization of a given parametrization $\cal M$, if the input surface is a  ruled surface.  More precisely, if $\cal V$ is a  rational  ruled surface, then there exists  a parametrization given in standard reduced form
\[{\cal P}(\ot)=(p_1(t_1)+t_2q_1(t_1), p_2(t_1)+t_2q_2(t_1), t_2)\in {\Bbb K}(\ot)^3 \]
(see equation given in~\ref{eq-standardreducedform}), where $p_j, q_j$ are given in Theorem~\ref{T-implicharac1}. Thus, we only have to check whether there exists   $(U, V)\in ({\Bbb K}(\ot)\setminus{\Bbb K})^2$ such that ${\cal P}(U, V)={\cal M}$. Observe that from this equality, we get that $V=m_3$.

 \para

 To start with, we prove the following theorem that is equivalent to Theorem~\ref{T-implicharac1} and Corollary~\ref{C-implicharac1}, but for the parametric case.

 \para

\begin{theorem}\label{T-paracharac1} A surface $\cal V$ defined by the parametrization   \[{\cal M}(\ot)=(m_1(\ot), m_2(\ot), m_3(\ot))\in {\Bbb K}(\ot)^3\]  is a rational ruled surface if and only if the following statements holds:
\begin{itemize}
\item[1.]  There exists two rational   plane curves ${\cal C}^{ij}$ and ${\cal C}^{k\ell}$ defined by a factor of $f_0^{ij}$ and $f_0^{k\ell}$, respectively,  for $ij\not=k\ell$ and $ij, k\ell\in \{12,23,13\}$ (see Section 2). Let $ij=12$, and $k\ell=23$, and let $\cP^{12}=(p_1,p_2)\in {\Bbb K}(t_1)^2$, $\cP^{23}=(\tilde{p}_1,\tilde{p}_2)\in {\Bbb K}(t_1)^2$ be rational  proper parametrizations of ${\cal C}^{12}$ and ${\cal C}^{23}$, respectively.

\item[2.] If $p_1\not=0$, there exists $({\cal L}, {\cal T})\in {\Bbb K}(\ot)\setminus{\Bbb K}$  such that one of the following statements holds:
\begin{itemize}
\item[2.1.]
     $p_1({\cal T})-m_3\frac{p_1({\cal T})}{\tilde{p}_2({\cal L})}-m_1= p_2({\cal T})+m_3\frac{\tilde{p}_1({\cal L})-p_2({\cal T})}{\tilde{p}_2({\cal L})}-m_2=0.$
  In this case, \[{\cal P}(\ot)=\left(p_1(S(t_1))-t_2\frac{p_1(S(t_1))}{\tilde{p}_2(R(t_1))}, p_2(S(t_1))+t_2\frac{\tilde{p}_1(R(t_1))-p_2(S(t_1))}{\tilde{p}_2(R(t_1))},t_2\right),\] is a  rational proper parametrization  in standard reduced form of $\cal V$, where $(R, S)\in ({\Bbb K}(t_1)\setminus{\Bbb K})^2$ is  a rational proper parametrization of the curve ${\cal C}_{N_1}$ that is defined parametrically by $({\cal L}, {\cal T})$.
 \item[2.2.] $p_1({\cal T})-m_3\frac{p_1({\cal T})}{\tilde{p}_2({\cal L})}-m_1=p_2({\cal T})-m_2=0.$
 In this case, \[{\cal P}(\ot)=\left(p_1(S(t_1))-t_2\frac{p_1(S(t_1))}{\tilde{p}_2(R(t_1))}, p_2(S(t_1)),t_2\right)\]
 is a rational proper parametrization  in standard reduced form of $\cal V$, where $(R, S)\in ({\Bbb K}(t_1)\setminus{\Bbb K})^2$  is a rational proper parametrization of the curve ${\cal C}_{N_2}$ that is defined parametrically by $({\cal L}, {\cal T})$.
 \end{itemize}

\item[3.] If $p_1=0$,  there exists ${\cal L}\in {\Bbb K}(\ot)\setminus{\{0\}}$, and ${\cal T}\in {\Bbb K}(\ot)\setminus{\Bbb K}$   such that one of the following statements holds:
\begin{itemize}
\item[3.1.]
$$\left\{\begin{array}{ll}
m_3\frac{q_1({\cal T})}{q_2({\cal T})}-m_1={\cal L}-m_3\frac{{\cal L}}{{q}_2({\cal T})}-m_2=0,&\begin{array}{l}  \mbox{if $\cP^{13}=(q_1,q_2),\,q_2\not=0$ is a}\\ \mbox{proper parametrization of  ${\cal C}^{13}$} \end{array}\\
\\
m_3{\cal T}-m_1=m_3{\cal L}-m_2=0,\,{\cal L}\not\in{\Bbb K}  &\begin{array}{l}  \mbox{if $\cP^{13}=(t_1,0)$ is a}\\ \mbox{ parametrization of   ${\cal C}^{13}$} \end{array}\\
\end{array}
\right.$$
  In this case,
 $$\left\{\begin{array}{ll}
{\cal P}(\ot)=\left(t_2\frac{q_1(S(t_1))}{q_2(S(t_1))},  R(t_1)-t_2\frac{R(t_1)}{{q}_2(S(t_1))},t_2\right),&\begin{array}{l}  \mbox{if $\cP^{13}=(q_1,q_2),\,q_2\not=0$ is a}\\ \mbox{proper parametrization of  ${\cal C}^{13}$} \end{array}\\
\\
{\cal P}(\ot)=\left(t_2S(t_1),  t_2R(t_1),t_2\right),\,\,R\not\in{\Bbb K}  &\begin{array}{l}  \mbox{if $\cP^{13}=(t_1,0)$ is a}\\ \mbox{ parametrization of   ${\cal C}^{13}$} \end{array}\\
\end{array}
\right.$$
   is a  rational proper parametrization in standard reduced form of $\cal V$,  where $(R, S)\in {\Bbb K}(t_1)^2,\,S\not\in{\Bbb K}$ is  a rational proper parametrization of the curves  ${\cal C}_{N_3}$ or  ${\cal C}_{N_4}$, respectively,  that are defined  parametrically by $({\cal L}, {\cal T})$.
\item[3.2.]   $m_3{\cal T}-m_1={\cal L}-m_2=0$.
  In this case, \[{\cal P}(\ot)=\left(t_2S(t_1),  R(t_1),t_2\right),\] is a  rational proper parametrization  in standard reduced form of $\cal V$,  where $(R, S)\in ({\Bbb K}(t_1)\setminus{\Bbb K})^2$ is  a rational proper parametrization of the curve ${\cal C}_{N_5}$ that is defined parametrically by $({\cal L}, {\cal T})$.
   \end{itemize}
 \end{itemize}
\end{theorem}

\vspace*{2mm}

\noindent {\bf Proof.} It is clear that if statements $1$ and $2$ (or $3$) hold, then  $\cal V$ is a   rational  ruled surface. Reciprocally, let $\cal V$ be a  rational  ruled surface. Then, statement 1 holds (see statement 1 in  Theorem~\ref{T-implicharac1}), and some of the statements, $2$ or $3$, of Theorem~\ref{T-implicharac1} holds. Let us assume that statement $2.1$ holds (one reasons similarly if a different statement holds). That is,
\[{\cal P}^*(t_1, t_2)=\left(p_1({S}^*(t_1))-t_2\frac{p_1({S}^*(t_1))}{\tilde{p}_2({R}^*(t_1))}, p_2({S}^*(t_1))+t_2\frac{\tilde{p}_1({R}^*(t_1))-p_2({S}^*(t_1))}{\tilde{p}_2({R}^*(t_1))},t_2\right) \]
 is a proper parametrization of $\cal V$, where  $({R}^*, {S}^*)\in ({\Bbb K}(t_1)\setminus{\Bbb K})^2$ is a rational proper parametrization of the curve ${\cal C}_{N_1}$ (see Corollary~\ref{C-implicharac1}). Since $\cal M$ is also a  parametrization of $\cal V$, we have that  there exists $(U, V)\in ({\Bbb K}(\ot)\setminus{\Bbb K})^2$ such that ${\cal P}^*(U, V)={\cal M}$. From this equality, we get that $V=m_3$, and
 \[p_1({S}^*(U))-m_3\frac{p_1({S}^*(U))}{\tilde{p}_2({R}^*(U))}=m_1,\quad  p_2({S}^*(U))+m_3\frac{\tilde{p}_1({R}^*(U))-p_2({S}^*(U))}{\tilde{p}_2({R}^*(U))}=m_2.\]
 That is,
     $$p_1({\cal T})-m_3\frac{p_1({\cal T})}{\tilde{p}_2({\cal L})}-m_1= p_2({\cal T})+m_3\frac{\tilde{p}_1({\cal L})-p_2({\cal T})}{\tilde{p}_2({\cal L})}-m_2=0,$$
     where $({\cal L}, {\cal T}):=({R}^*(U), {S}^*(U))\in ({\Bbb K}(\ot)\setminus{\Bbb K})^2$.
    Observe that since $({R}^*, {S}^*)$ is a rational parametrization of   ${\cal C}_{N_1}$, then  $({\cal L}, {\cal T})$ also parametrizes  ${\cal C}_{N_1}$.\\
    Now, we consider $(R(t_1), S(t_1))\in ({\Bbb K}(t_1)\setminus{\Bbb K})^2$ a new rational proper parametrization of ${\cal C}_{N_1}$, and
   \[{\cal P}(t_1, t_2)=\left(p_1(S(t_1))-t_2\frac{p_1(S(t_1))}{\tilde{p}_2(R(t_1))}, p_2(S(t_1))+t_2\frac{\tilde{p}_1(R(t_1))-p_2(S(t_1))}{\tilde{p}_2(R(t_1))},t_2\right).\]
 Since  $({R}^*, {S}^*)$  and  $(R, S)$ are both  rational proper   parametrizations of  ${\cal C}_{N_1}$,  we have that there exists $r\in {\Bbb K}(t_1)\setminus{\Bbb K}$,\,$\deg(r)=1,$ such that $({R}^*, {S}^*)=({R}(r), {S}(r))$. Then, ${\cal P}(r(t_1), t_2)={\cal P}^*(t_1, t_2)$ which implies that   $\cal P$ is  a  rational proper parametrization of $\cal V$ (note that $(r(t_1), t_2)$ and ${\cal P}^*(\ot)$ are both proper, and thus ${\cal P}(\ot)$ is also proper).
 \qed

\para

From Theorem~\ref{T-paracharac1}, we obtain the following algorithm that decides whether a rational surface defined parametrically by a rational parametrization $\cal M$ is ruled, and in the affirmative case, it computes a rational proper reparametrization.

\para

\noindent
\fbox{{\sf Algorithm Reparametrization of a Ruled Surface.}}
\begin{itemize}
\item {\bf Input:} A surface $\cal V$ defined by a  rational parametrization
 \[{\cal M}(\ot)=(m_1(\ot), m_2(\ot), m_3(\ot))\in {\Bbb K}(\ot)^3.\]
\item {\bf Output:} the message ``{\tt $\cal V$ is not a ruled surface}" or
a proper parametrization ${\cal P}$ of ``{\tt the ruled surface $\cal V$}"
\end{itemize}
 \begin{itemize}
  \item[1.] If $m_1\in {\Bbb K}$ (similarly if $m_j\in {\Bbb K}$ for some $j=2, 3$), {\sc Return} ${\cal P}(\ot)=\left(m_1,  t_1, t_2\right)$
    ``{\tt  is a proper parametrization of the ruled surface  ${\cal V}$  in stan\-dard re\-du\-ced form}". Otherwise, go to Step 2.
    \item[2.] Compute the polynomials  $H_i(\ot,\oh)=\numer(m_{i}(\ot)-m_{i}(\oh))$,
 where $\oh=(h_1,h_2)$, and $i\in\{1,2,3\}$. Check whether $\gcd(H_i,H_j)= 1$, for $i,j \in\{1,2,3\}$ and $i<j$. In the affirmative case,   go to Step 3. Otherwise, if $\gcd(H_1,H_2)\neq 1$ (similarly if $\gcd(H_1,H_3)\neq 1$ or $\gcd(H_2,H_3)\neq 1$) do:
 \begin{itemize}
 \item[2.1.] Consider   $a\in \mathbb  K$ such that $(m_1, m_2)(a,t_2)\not\in {\Bbb K}^2$.
 \item[2.2.] Compute \[f(x_1,x_2)^{r}=\Resultant_{t_2}(G_1(a,t_2,x_2), G_2(a,t_2,x_2)),\quad \mbox{where}\quad r\in {\Bbb K}, \quad \mbox{and}\]
$G_i(\ot, x_i)=\numer(m_{i}(\ot)-x_{i}),\,i=1,2$ (see Theorem 8 in \cite{sonia08}).
 \item[2.3.] Determine a proper parametrization $(p(t_1), q(t_1))\in {\Bbb K}(t_1)^2$ of the plane curve defined by the equation $f(x_1, x_2)=0$.
  \item [2.4.] {\sc Return} ${\cal P}(\ot)=\left(p(t_1),q(t_1), t_2\right),$
    ``{\tt  is a proper parametrization of the ruled surface  ${\cal V}$  in stan\-dard re\-du\-ced form}".
 \end{itemize}

   \item[3.] Compute the polynomials $f_{0}^{ij}(x_i, x_j)$ (apply Theorem~\ref{T-impli}), and check whether there exist two rational  plane curves ${\cal C}^{ij}$ and ${\cal C}^{k\ell}$ defined by a factor of $f_0^{ij}$ and $f_0^{k\ell}$, respectively,  for $ij\not=k\ell$, and $ij, k\ell\in \{12,23,13\}$ . In the affirmative case,  we assume that $ij=12$, and $k\ell=23$ (see Remark~\ref{R-ij}). Otherwise, {\sc Return} ``{\tt $\cal V$ is not a ruled surface}".
         \item[4.] Compute $\cP^{12}=(p_1,p_2)\in {\Bbb K}(t_1)^2$, and $\cP^{23}=(\tilde{p}_1,\tilde{p}_2)\in {\Bbb K}(t_1)^2$ rational proper parametrizations of ${\cal C}^{12}$ and ${\cal C}^{23}$, respectively. For this purpose, apply for instance the results in Sections 4.7 and 4.8 in \cite{libro}.  If $p_1\not=0$ go to Step 5. Otherwise, go to Step 7.
     \item[5.] Check whether there exists $({\cal L}, {\cal T})\in ({\Bbb K}(\ot)\setminus{\Bbb K})^2$ such that
     \[p_1({\cal T})-m_3\frac{p_1({\cal T})}{\tilde{p}_2({\cal L})}=m_1,\qquad p_2({\cal T})+m_3\frac{\tilde{p}_1({\cal L})-p_2({\cal T})}{\tilde{p}_2({\cal L})}=m_2.\]
     In the affirmative case, compute $(R, S)\in ({\Bbb K}(t_1)\setminus{\Bbb K})^2$   a rational proper parametrization of the curve ${\cal C}_{N_1}$ defined by $({\cal L}, {\cal T})$, and
{\sc Return} \[{\cal P}(\ot)=\left(p_1(S(t_1))-t_2\frac{p_1(S(t_1))}{\tilde{p}_2(R(t_1))}, p_2(S(t_1))+t_2\frac{\tilde{p}_1(R(t_1))-p_2(S(t_1))}{\tilde{p}_2(R(t_1))},t_2\right),\]
    ``{\tt  is a proper parametrization of the ruled surface  ${\cal V}$  in stan\-dard re\-du\-ced\- form}". Otherwise, go to Step 6.
\item[6.]  Check whether there exists $({\cal L}, {\cal T})\in ({\Bbb K}(\ot)\setminus{\Bbb K})^2$ such that
     \[p_1({\cal T})-m_3\frac{p_1({\cal T})}{\tilde{p}_2({\cal L})}=m_1,\qquad  p_2({\cal T})=m_2.\]
     In the affirmative case, compute $(R, S)\in ({\Bbb K}(t_1)\setminus{\Bbb K})^2$   a rational proper parametrization of the curve ${\cal C}_{N_2}$ defined by  $({\cal L}, {\cal T})$, and
  {\sc Return}
\[{\cal P}(\ot)=\left(p_1(S(t_1))-t_2\frac{p_1(S(t_1))}{\tilde{p}_2(R(t_1))}, p_2(S(t_1)),t_2\right)\]
   ``{\tt  is a proper parametrization of the ruled surface  ${\cal V}$  in  stan\-dard re\-du\-ced\- form}".  Otherwise, go to Step 3, and consider different rational components and apply again the algorithm. If there have no more rational components, {\sc Return} ``{\tt $\cal V$ is not a ruled surface}".
   \item[7.]  Check whether the plane curve ${\cal C}^{13}$ is rational. In the affirmative case, compute   $\cP^{13}=(q_1,q_2)\in {\Bbb K}(t_1)^2$   a rational  proper parametrization  of ${\cal C}^{13}$.  Otherwise, go to Step 9.
            \item[8.]
            \begin{itemize}
                \item[8.1.]         If $q_2\not=0$, check whether there exists $({\cal L}, {\cal T})\in {\Bbb K}(\ot)^2,\,{\cal T}\not\in{\Bbb K}$ such that
     \[m_3\frac{q_1({\cal T})}{q_2({\cal T})}=m_1,\qquad   {\cal L}-m_3\frac{{\cal L}}{{q}_2({\cal T})}=m_2.\]
     In the affirmative case, compute $(R, S)\in {\Bbb K}(t_1)^2,\,\,S\not\in {\Bbb K}$   a rational proper parametrization of the curve ${\cal C}_{N_3}$ defined by  $({\cal L}, {\cal T})$,       and  {\sc Return} \[{\cal P}(\ot)=\left(t_2\frac{q_1(S(t_1))}{q_2(S(t_1))},  R(t_1)-t_2\frac{R(t_1)}{{q}_2(S(t_1))},t_2\right),\]
    ``{\tt  is a proper parametrization of the ruled surface  ${\cal V}$  in stan\-dard re\-du\-ced form}". Otherwise, go to Step 3, and consider different rational components and apply again the algorithm. If there have no more rational components, {\sc Return} ``{\tt $\cal V$ is not a ruled surface}".
                    \item[8.2.]      If $q_2=0$, let ${\cal L}=m_2/m_3$, and ${\cal T}=m_1/m_3$, and compute $(R, S)\in ({\Bbb K}(t_1)\setminus{\Bbb K})^2$   a rational proper parametrization of the curve ${\cal C}_{N_4}$ defined by  $({\cal L}, {\cal T})$, and   {\sc Return} \[{\cal P}(\ot)=\left(t_2S(t_1),  t_2R(t_1),t_2\right),\]
    ``{\tt  is a proper parametrization of the ruled surface  ${\cal V}$  in stan\-dard re\-du\-ced form}".  Otherwise, go to Step 3, and consider different rational components and apply again the algorithm. If there have no more rational components, {\sc Return} ``{\tt $\cal V$ is not a ruled surface}".
    %check whether there exists $(x,y)\in ({\Bbb K}(\ot)\setminus{\Bbb K})^2$ such that      \[m_3y=m_1,\qquad   m_3x=m_2.\]      In the affirmative case, compute
      \end{itemize}

                   \item[9.] Let ${\cal L}=m_2$, and ${\cal T}=m_1/m_3$, and compute $(R, S)\in ({\Bbb K}(t_1)\setminus{\Bbb K})^2$   a rational proper parametrization of the curve ${\cal C}_{N_5}$  defined by $({\cal L}, {\cal T})$.    {\sc Return} \[{\cal P}(\ot)=\left(t_2S(t_1),  R(t_1),t_2\right),\]
    ``{\tt  is a proper parametrization of the ruled surface  ${\cal V}$  in stan\-dard re\-du\-ced form}". Otherwise, go to Step 3, and consider different rational components and apply again the algorithm. If there have no more rational components, {\sc Return} ``{\tt $\cal V$ is not a ruled surface}".
\end{itemize}

%\para

 \begin{remark}\label{R-alg2} \begin{enumerate}\item
    Remark~\ref{R-Alg2} can be stated similarly in this new situation to   the rational parametrizations $\cP^{12}, \cP^{23}, \cP^{13}$, $(R,S)$, and the output parametrization $\cal P$.
      \item To find $({\cal L}, {\cal T})\in ({\Bbb K}(\ot)\setminus{\Bbb K})^2$, we should solve the system to get a rational solution (note we have two equations with two unknowns).
For this purpose,  for instance one may use univariate resultants. Once $({\cal L}, {\cal T})$ is determined, one computes the implicit equation of the rational  plane  curve ${\cal C}_{N_i}$ defined by $({\cal L}, {\cal T})$ as the square free part of
         \[\content_{t_2}(\resultant_{t_1}(\numer({\cal L}-x_1), \numer({\cal T}-x_2)))\in {\Bbb K}[x_1,x_2]\]
     (see Section 4.5 in \cite{libro}). Afterwards, we parametrize  ${\cal C}_{N_i}$ by applying for instance the results in Sections 4.7 and 4.8 in \cite{libro}.
     \item Note that in this case, we can not apply Theorem~\ref{T-implicharac2}.  More precisely, if $\cal V$ is a  rational  ruled surface, by  Theorem~\ref{T-implicharac2}, there exists  a parametrization of the form
  \[{\cal P}^*(\ot)=(p_1(S^*(t_1))+t_2q^*_1(t_1), p_2(S^*(t_1))+t_2q^*_2(t_1), t_2)\in {\Bbb K}(\ot)^3\] where  $(q^*_1, q^*_2, S^*)$ is a rational parametrization of a space curve  ${\cal D}$ (see Corollary~\ref{C-implicharac2}). Reasoning as in Theorem~\ref{T-paracharac1}, we have that   ${\cal P}^*(U, V)={\cal M}$, where $(U, V)\in ({\Bbb K}(\ot)\setminus{\Bbb K})^2$.  From this equality, we get that $V=m_3$, and
 \[p_1(S^*(U))+m_3q^*_1(U)=m_1,\quad  p_2({S}^*(U))+m_3{q}^*_2(U)=m_2.\]
 That is,
     $$p_1({\cal T})+m_3{\cal L}_1-m_1=p_2({\cal T})+t_2{\cal L}_2-m_2=0, $$ where $({\cal L}_1, {\cal L}_2, {\cal T}):=({q^*_1}(U),{q^*_2}(U), {S}^*(U))\in ({\Bbb K}(\ot)\setminus{\Bbb K})^3$. Observe that we have two equations, and  three unknowns ${\cal L}_1, {\cal L}_2, {\cal T}$. So, we have a consistent independent system.
\end{enumerate}
\end{remark}

\para

In the following example, we illustrate the performance of  {\sf Algorithm Parametrization of a Ruled Surface}.

\para

\begin{example} Consider the surface $\cal V$ defined by the parametrization
\[{\cal M}(\ot)=(m_1, m_2, m_3)=\left(-\frac{2  t_2^4  t_1+10  t_2^2  t_1^3+5  t_2  t_1^4-7  t_2^3  t_1^2-5  t_1^3-9  t_1^2  t_2+7  t_1  t_2^2- t_2^3}{ t_2 (- t_1^2-2  t_1  t_2+ t_2^2) ( t_2^2+ t_1^2)},\right.\]
\[\left. \frac{- t_2 (-14  t_1^2  t_2^2+4  t_1^4+4  t_2^3  t_1-14  t_1^3  t_2+9  t_1^2+18  t_1  t_2-9  t_2^2}{(- t_1^2-2  t_1  t_2+ t_2^2) ( t_2^2+ t_1^2)  t_1}, \frac{ t_1  t_2-1}{ t_2^2+ t_1^2}\right)\in {\Bbb R}(\ot)^3.\]
Let us apply {\sf Algorithm Reparametrization of a Ruled Surface}. For this purpose, we first observe that $\cal V$  is neither a cylinder nor a plane (see Steps 1 and 2 of the algorithm). In Step 3 of the algorithm, we apply Theorem~\ref{T-impli}, and we compute the polynomials
 $f_{0}^{ij}(x_i, x_j)$.  We get that
$$f_0^{12}(x_1,x_2)=49x_1^2-20x_1-2x_1x_2+5x_2-x_2^2,$$
$$f_0^{23}(x_2,x_3)= x_2^3-14 x_2^2-25x_3 x_2^2+1396 x_3 x_2-1120x_3-4237 x_3^2 x_2-5190 x_3^2-48915 x_3^3,$$
%$$f_0^{13}(x_1,x_3)=441 x_3 x_1^2-196 x_1^2+80 x_1+5 x_3 x_1+224 x_3+6750x_3^2 x_1+1038 x_3^2+9783 x_3^3.$$
define implicitly two rational  plane curves   ${\cal C}^{12}$ and ${\cal C}^{23}$. Thus, in Step 4,  we compute $$\cP^{12}=(p_1,p_2)=\left(\frac{-\sqrt{2}t_1(-5+t_1)}{5(4t_1-10+5\sqrt{2})}, \frac{(50+5\sqrt{2})t_1(-20+100\sqrt{2}+49t_1)}{1225(4t_1-10+5\sqrt{2})}\right)\in {\Bbb R}(t_1)^2,$$ and $$\cP^{23}=(\tilde{p}_1(t_1),\tilde{p}_2(t_1))=\left(\frac{2 (-378367 t_1^2+10410900 t_1-142098075+4102 t_1^3)}{1241(t_1^3-25 t_1^2-4237 t_1-48915)}, \right.$$ $$ \left.\frac{2(20322550-513355 t_1-28 t_1^2+49 t_1^3)}{1241(t_1^3-25 t_1^2-4237 t_1-48915)}\right)\in {\Bbb R}(t_1)^2$$  rational proper parametrizations of ${\cal C}^{12}$ and ${\cal C}^{23}$, respectively. Hence, we go to  Step 5 of the algorithm, and we check whether there exists $({\cal L},{\cal T})\in ({\Bbb C}(\ot)\setminus{\Bbb C})^2$ such that
     \[p_1({\cal T})-m_3\frac{p_1({\cal T})}{\tilde{p}_2({\cal L})}=m_1,\qquad p_2({\cal T})+m_3\frac{\tilde{p}_1({\cal L})-p_2({\cal T})}{\tilde{p}_2({\cal L})}=m_2.\]
       We obtain
  \[ {\cal L}(\ot)= \frac{415t_1-236t_2}{2t_2+7t_1},\quad {\cal T}(\ot)= \frac{-5 (-1+5\sqrt{2})t_2}{9t_1+4t_1\sqrt{2}-5t_2\sqrt{2}+t_2}.\]
     $({\cal L}, {\cal T})$ parametrizes the rational  plane curve ${\cal C}_{N_1}$ defined implicitly by the equation
     \[-315x_1+18675+70\sqrt{2}x_1-4150\sqrt{2}+2482x_2\sqrt{2}+73x_2x_1-2555x_2=0 \]
     (see statement~2 in Remark~\ref{R-alg2}). We compute  a rational proper parametrization of ${\cal C}_{N_1}$, and we get $$(R(t_1), S(t_1))=\left(t_1,  \frac{-5(-9+2\sqrt{2})(7t_1-415)}{73(34\sqrt{2}+t_1-35)}\right)\in {\Bbb R}(t_1)^2.$$ Therefore, we
{\sc Return} the proper parametrization of the ruled surface  ${\cal V}$ given by
 \[{\cal P}(\ot)=\left(p_1(S(t_1))-t_2\frac{p_1(S(t_1))}{\tilde{p}_2(R(t_1))}, p_2(S(t_1))+t_2\frac{\tilde{p}_1(R(t_1))-p_2(S(t_1))}{\tilde{p}_2(R(t_1))},t_2\right)=\]\[=\left(\frac{q_{11}(\ot)}{q_{12}(\ot)},\frac{q_{21}(\ot)}{q_{22}(\ot)},t_2\right)\in {\Bbb R}(\ot)^3,\]
 where\\

 \noindent
 $q_{11}=\sqrt{2} (-9+2 \sqrt{2}) (5+7 \sqrt{2}) (-40645100+1026710 t_1+56 t_1^2-98 t_1^3+1241 t_1^3 t_2-31025 t_1^2 t_2-5258117 t_1 t_2-60703515 t_2),$\\

 \noindent
 $q_{12}=73(7 t_1-415) (-4199 \sqrt{2}+106 t_1-4866+17 t_1 \sqrt{2}) (34 \sqrt{2}+t_1-35),$\\

 \noindent
 $q_{21}=78183 t_2 t_1^3 \sqrt{2}+63812 t_1^3+487494 t_1^3 t_2+10234 t_1^3 \sqrt{2}+331704 t_1^2 \sqrt{2}-10107945 t_1^2 t_2 \sqrt{2}-63026010 t_1^2 t_2+2068272 t_1^2-87398734 t_1 \sqrt{2}-544956812 t_1+239474529 t_2 t_1 \sqrt{2}+1493194122 t_1 t_2+11741634840+5038391745 t_2 \sqrt{2}+31415854410 t_2+1883092380 \sqrt{2},$\\

 \noindent
 $q_{22}=146(34 \sqrt{2}+t_1-35) (-4199 \sqrt{2}+106 t_1-4866+17 t_1 \sqrt{2}) (118+t_1).$\\

    \noindent Observe that since   $\cP^{12}, \cP^{23}$ and $(R,S)$   have coefficients in $\Bbb R$, then the output parametrization $\cal P$  also has coefficients in  $\Bbb R$ (see statement~1 in Remark~\ref{R-alg2}).
      \end{example}

%----------------------------------------------------------------
\section{Conclusion}
The parametrization of an implicit surface is a basic problem in algebraic geometry. In this paper, we focus on the problem of  rational  ruled surface, since the ruled surface is an important modeling surface. By the linearity of one parameter in the standard form and the birational parameter transformation, we can get a simple expression which can be projected as a planar curve. Therefore we reduce the problem to that of curve parametrization. The algorithms to determine and parameterize the implicit  rational  ruled surfaces are then proposed. We also have considered the determination and reparametrization for the parametric ruled surfaces not being in the standard form. More precisely, we can distinguish whether a given rational parametrization (not necessarily proper) defines a ruled surface, and in the affirmative case, we may reparameterize properly to the standard form.

\para

Besides the ruled surface, there are some other basic modeling surfaces such as sphere-swept surfaces and cyclides. They have special geometric features for modeling design. And these features are also reflected in the algebraic expressions. According to the well investigation, one can find some algorithms to determine the types of surface from a given algebraic surface, further, find a parametrization. As the further work, we would like given more discussions for other modeling surfaces.

\end{document}